\documentclass[10pt,twocolumn,reqno, aps,prx,superscriptaddress]{revtex4-2}%
\usepackage{eurosym}
\usepackage{amsmath}
\usepackage[pdftex]{graphicx}
\usepackage{float}
\usepackage{color}
\usepackage{rotating}
\usepackage[breaklinks=true,colorlinks,citecolor=blue,linkcolor=blue,urlcolor=blue]%
{hyperref}
\usepackage{graphicx}
\usepackage{natbib}
\usepackage{dcolumn}
\usepackage{bm}
\usepackage{hyperref}
\usepackage[caption=false]{subfig}
\usepackage{epstopdf}
\usepackage{ulem}
\usepackage{amsfonts}
\usepackage{amssymb}%
\providecommand{\U}[1]{\protect\rule{.1in}{.1in}}
\providecommand{\U}[1]{\protect\rule{.1in}{.1in}}
\providecommand{\U}[1]{\protect\rule{.1in}{.1in}}
\providecommand{\U}[1]{\protect\rule{.1in}{.1in}}
\begin{document}

\title{Effect of nonlinear magnon interactions on the stochastic magnetization switching}
\author{Mehrdad Elyasi}
\affiliation{Advanced Institute for Materials Research, Tohoku University, Sendai 980-8577, Japan}
\affiliation{Center for Science and Innovation in Spintronics, Tohoku University, 2-1-1 Katahira, Sendai 980-8577, Japan.}
\author{Shun Kanai}
\affiliation{Laboratory for Nanoelectronics and Spintronics, Research Institute of Electrical Communication, Tohoku University, 2-1-1 Katahira, Sendai 980-8577, Japan}
\affiliation{Graduate School of Engineering, Tohoku University, 6-6 Aramaki Aza Aoba, Sendai 980-8579, Japan. }
\affiliation{Advanced Institute for Materials Research, Tohoku University, Sendai 980-8577, Japan}
\affiliation{Center for Science and Innovation in Spintronics, Tohoku University, 2-1-1 Katahira, Sendai 980-8577, Japan.}
\affiliation{PRESTO, Japan Science and Technology Agency (JST), Kawaguchi 332-0012, Japan. }
\affiliation{Division for the Establishment of Frontier Sciences of Organization for Advanced Studies at Tohoku University, Tohoku University, Sendai 980-8577, Japan.}
\affiliation{National Institutes for Quantum Science and Technology.}
\author{Hideo Ohno}
\affiliation{Center for Science and Innovation in Spintronics, Tohoku University, 2-1-1 Katahira, Sendai 980-8577, Japan.}
\affiliation{Center for Innovative Integrated Electronic Systems, Tohoku University, 468-1 Aramaki Aza Aoba, Sendai 980-8572, Japan.}
\author{Shunsuke Fukami}
\affiliation{Laboratory for Nanoelectronics and Spintronics, Research Institute of Electrical Communication, Tohoku University, 2-1-1 Katahira, Sendai 980-8577, Japan}
\affiliation{Graduate School of Engineering, Tohoku University, 6-6 Aramaki Aza Aoba, Sendai 980-8579, Japan. }
\affiliation{Advanced Institute for Materials Research, Tohoku University, Sendai 980-8577, Japan}
\affiliation{Center for Science and Innovation in Spintronics, Tohoku University, 2-1-1 Katahira, Sendai 980-8577, Japan.}
\affiliation{Center for Innovative Integrated Electronic Systems, Tohoku University, 468-1 Aramaki Aza Aoba, Sendai 980-8572, Japan.}
\affiliation{Inamori Research Institute of Science, Shijo, Shimogyo-ku, Kyoto 600-8411, Japan.}
\author{Gerrit E. W. Bauer}
\affiliation{Advanced Institute for Materials Research, Tohoku University, Sendai 980-8577, Japan}
\affiliation{Center for Science and Innovation in Spintronics, Tohoku University, 2-1-1 Katahira, Sendai 980-8577, Japan.}
\affiliation{Kavli Institute for Theoretical Sciences, University of the Chinese Academy of Sciences, Beijing 10090, China}
\affiliation{Zernike Institute for Advanced Materials, University of Groningen, 9747 AG Groningen, Netherlands}

\begin{abstract}
Telegraph noise caused by frequent switching of the magnetization in small
magnetic devices has become a useful resource for probabilistic computing.
Conventional theories have been based on a linearization of the fluctuations
at the extrema of the magnetic free energy. We show theoretically that the
non-linearities, specifically four-magnon scatterings, reduce the equilibrium fluctuation amplitude of the
magnetization as well as the switching frequencies between local minima via
the decay of the homogeneous Kittel mode into two spin waves with opposite
momenta. Selectively suppressing the effective temperature of the finite-$k$ spin waves, or reducing the radius
of a thin magnetic disk enhance the switching frequency and improve
performance of magnetic tunnel junctions in probabilistic computing applications.

\end{abstract}
\maketitle
\date{\today}

\section{\label{Introduction} Introduction}

Fluctuations in magnetic materials are unwanted in applications such as data
storage and communication, but are also essential for probabilistic computing
or quantum information with magnetic devices. Research on magnetic noise
focuses on either (i) the large fluctuations associated with equilibrium
random telegraph noise
(RTN)\cite{Neel1949,Brown1963,Brown1979,Hayakawa2021,Safranski2021} and
quantum tunneling of the magnetization \cite{Chudnovsky1988,Awschalom1992}, as
well as the Barkhausen noise due to moving magnetic textures
\cite{Barkhausen1919,Spasojevic1996}, or (ii) the small fluctuations around
the equilibrium magnetization, e.g. thermal noise of spin waves including the
uniform (Kittel) mode
\cite{Lachance2020,Elyasi2020,Yuan2022,Rameshti2022,Chumak2022,Hioki2021,Makiuchi2024}%
. Under nonlinear (parametric) excitation, multi-stability of the
magnetization can be established in the dynamical phase space that allows to
manipulate the RTN and quantum tunneling between attractors
\cite{Makiuchi2021,Elyasi2022}. On the other hand, the magnetization in a
magnet is generally bistable, which becomes a source of RTN
\cite{Brown1979,Hayakawa2021,Endean2014,Talatchian2021} and can be a resource
in probabilistic computing
\cite{Camsari2017,Borders2019,Vodenicarevic2017,Mizrahi2018,Kaiser2022,Singh2024,Si2024}%
. While advances in fabricating dedicated magnetic tunnel junctions (MTJ)
\cite{Ikeda2010,Watanabe2018,Jinnai2020} already led to prototype
devices \cite{Borders2019,Hayakawa2021,Funatsu2022,Singh2024,Si2024} that could solve specific problems such as combinatorial optimization,
the computing speed is still an issue.

N\'{e}el and Brown \cite{Neel1949,Brown1963,Brown1979} formulated the RTN in a
ferromagnetic particle based on Kramers escape theory \cite{Kramers1940}
applied to the macrospin. In sufficiently large particles, thermally activated
magnetization reversal may also occur through domain wall nucleation and
motion
\cite{Braun1993,Braun1994,Braun1994_1,Braun1994_2,Krause2009,Bessarab2013,Sala2023}%
. Braun's \cite{Braun1993,Braun1994,Braun1994_1,Braun1994_2} theory and that
of others \cite{Bessarab2013} linearize the fluctuations around stable points
of the dynamics and therefore do not capture magnon interactions that emerge
from the nonlinearities at large fluctuations.

Phenomenologically, RTN is characterized by a stochastic switching frequency
$f_{s}$ that must be high for probabilistic computing applications
\cite{Hayakawa2021,Kanai2021,Safranski2021}. The Arrhenius law $f_{s}%
=f_{0}e^{-E_{B}/{k_{B}T}}$ holds for any thermally activated reaction
\cite{Arrhenius1889}, where $f_{0}$ is the attempt frequency, $k_{B}$ is the
Boltzmann constant, $E_{B}$ is a barrier energy, and $T$ is the temperature.
We recently found experimentally that the attempt time $\tau_{0}=1/f_{0}$ in
magnetic tunnel junctions is larger than expected for a macrospin model
\cite{Kanai2023,Kaneko2024}. Nonuniform switching via domain walls would
explain an increase rather than a decrease of $f_{s}$
\cite{Braun1993,Braun1994_1,Krause2009,Bessarab2013} and therefore cannot
explain the observations. Here, we demonstrate that the decay of the Kittel
macrospin into spin waves with finite linear momentum can explain the observed
brake on the stochastic switching.

In Sec. \ref{eqb_dist}, we show how magnon interactions affect the equilibrium
amplitude distribution of the spatially uniform magnon (Kittel mode). In Sec.
\ref{model}, we introduce the master equation for the Kittel magnon mode
coupled to a pair of spin waves. In Sec. \ref{analytic}, we analyze the
effects of magnon interactions on the distribution function and the effective
magnon occupation numbers. In Sec. \ref{results}, we show numerical results
for the equilibrium magnon numbers and distribution functions. In Sec.
\ref{RTN}, we focus on the effect of nonlinear magnon interactions on the RTN.
In Sec. \ref{past}, we briefly review different approaches to calculate
$f_{s}$, viz. exact calculations for the macrospin model and theories based on
linearization around the free energy extrema in the presence of domain walls.
In Sec. \ref{model1}, we introduce a simplified model for an easy numerical
analysis of the macrospin RTN including magnon interactions. In Sec.
\ref{results1}, we discuss a phenomenological analysis and the full numerical
model for the effect of nonlinear interactions on the RTN.

\section{\label{eqb_dist} Equilibrium distribution of interacting magnons}

The thermal switching without external drive corresponds to rare large
fluctuations in the equilibrium. According to the fluctuation dissipation
theorem these can be expressed in terms of the linear response to a
temperature or magnon density difference between different modes that induce
spin and heat currents between them. We first focus on the equilibrium
fluctuations of the Kittel and finite-$k$ spin wave modes around a stable
magnetization direction that form the input for Kramers's and Langer's
theories for the switching frequency $f_{s}$ in Sec. \ref{RTN}. The free layer
of state-of-the-art MTJs contain a disk of ultrathin magnetic films with
thickness of a few nanometer and lateral dimensions of a few tens of
nanometer. The exchange interaction then dominates the dispersion and shifts
the nonuniform modes to energies above the ferromagnetic resonance of the
lowest (homogeneous) Kittel mode.

According to the fluctuation dissipation theorem, we may approach the
stochastics of a strictly equilibrium system in terms of the response to a
thermodynamic force such as a temperature gradient. The latter can also be
intentionally studied for example by mode-selective active cooling
\cite{Aspelmeyer2014}. The latter selectivity would be easier to achieve for
small magnets because the magnon spectrum is discrete and finite-$k$ magnons
are far detuned from the Kittel mode. In the following we therefore consider
reservoirs of different magnon modes that may be at different temperatures.

\subsection{\label{model} Model}

The magnon Hamiltonian of the Kittel mode and spin waves up to the fourth
order in the Holstein-Primakoff expansion reads
\cite{Suhl1957,Krivosik2010,Elyasi2020,Elyasi2022}
\begin{align}
\mathcal{H}  &  =\sum_{\vec{k}}\omega_{\vec{k}}c_{\vec{k}}^{\dag}c_{\vec{k}%
}+\mathcal{H}^{(Suhl)}+\mathcal{H}^{(SK)}+\mathcal{H}^{(CK)},\label{eq1}\\
&  \mathcal{H}^{(Suhl)}=\nonumber\\
&  \sum_{\vec{k},\vec{k}_{1},\vec{k}_{2}}\mathcal{D}_{\vec{k},\vec{k}_{1}%
,\vec{k}_{2},\vec{k}+\vec{k}_{1}-\vec{k}_{2}}c_{\vec{k}}^{\dag}c_{\vec{k}_{1}%
}^{\dag}c_{\vec{k}_{2}}c_{\vec{k}+\vec{k}_{1}-\vec{k}_{2}}+H.c.,\label{eq2}\\
&  \mathcal{H}^{(SK)}=\sum_{\vec{k}}\mathcal{D}_{\vec{k},\vec{k},\vec{k}%
,\vec{k}}c_{\vec{k}}^{\dag}c_{\vec{k}}c_{\vec{k}}^{\dag}c_{\vec{k}%
},\label{eq3}\\
&  \mathcal{H}^{(CK)}=\sum_{\vec{k},\vec{k}_{1}}\mathcal{D}_{\vec{k},\vec
{k}_{1},\vec{k},\vec{k}_{1}}c_{\vec{k}}^{\dag}c_{\vec{k}}c_{\vec{k}_{1}}%
^{\dag}c_{\vec{k}_{1}}(1-\delta_{\vec{k},\vec{k}_{1}}), \label{eq4}%
\end{align}
where $\mathcal{D}_{\vec{k},\vec{k}_{1},\vec{k}_{2},\vec{k}_{3}}$ are the
strengths of the interaction $c_{\vec{k}}^{\dag}c_{\vec{k}_{1}}^{\dag}%
c_{\vec{k}_{2}}c_{\vec{k}_{3}}$ that depend on material parameters, sample
geometry, and magnetic field strength and direction. We assume that the magnet
is so thin that three magnon scattering is not resonant, i.e. the Kittel mode
frequency $\omega_{0}<\omega_{\vec{k}\neq0}$, and only renormalizes the four
magnon scattering amplitudes \cite{Krivosik2010}.

The master equation governing the density matrix equation of motion (EOM)
\cite{Carmichael1,Walls2008},
\[
\dot{\rho}=-i[\mathcal{H},\rho]+\sum_{\vec{k}}\xi_{\vec{k}}L_{\vec{k}}%
^{(L)}[\rho],
\]
where $\xi_{\vec{k}}=\alpha_{G}\omega_{\vec{k}}$ is the dissipation rate of
the magnon mode with wave vector $\vec{k}$, $\alpha_{G}$ is the Gilbert
damping assumed to be the same for all wave vectors, while the Lindblad
dissipation operator
\begin{align}
L_{\vec{k}}^{(L)} &  =(\bar{n}_{\vec{k}}+1)(2c_{\vec{k}}\rho c_{\vec{k}}%
^{\dag}-c_{\vec{k}}^{\dag}c_{\vec{k}}\rho-\rho c_{\vec{k}}^{\dag}c_{\vec{k}%
})+\nonumber\\
&  \bar{n}_{\vec{k}}(2c_{\vec{k}}^{\dag}\rho c_{\vec{k}}-c_{\vec{k}}c_{\vec
{k}}^{\dag}\rho-\rho c_{\vec{k}}c_{\vec{k}}^{\dag}),\label{eq6}%
\end{align}
with $\bar{n}_{\vec{k}}=\left[  \exp\left(  {\hbar\omega_{\vec{k}}%
/k_{B}T_{\vec{k}}}\right)  -1\right]  ^{-1},$ is the average number of thermal
bosons thermalized by baths at possibly different equilibrium temperatures
$T_{\vec{k}}$. The different bath temperatures model
selectively heated or cooled magnon modes, that bring the system into a
non-equilibrium state.

\begin{figure}[ptb]
\includegraphics[width=0.5\textwidth]{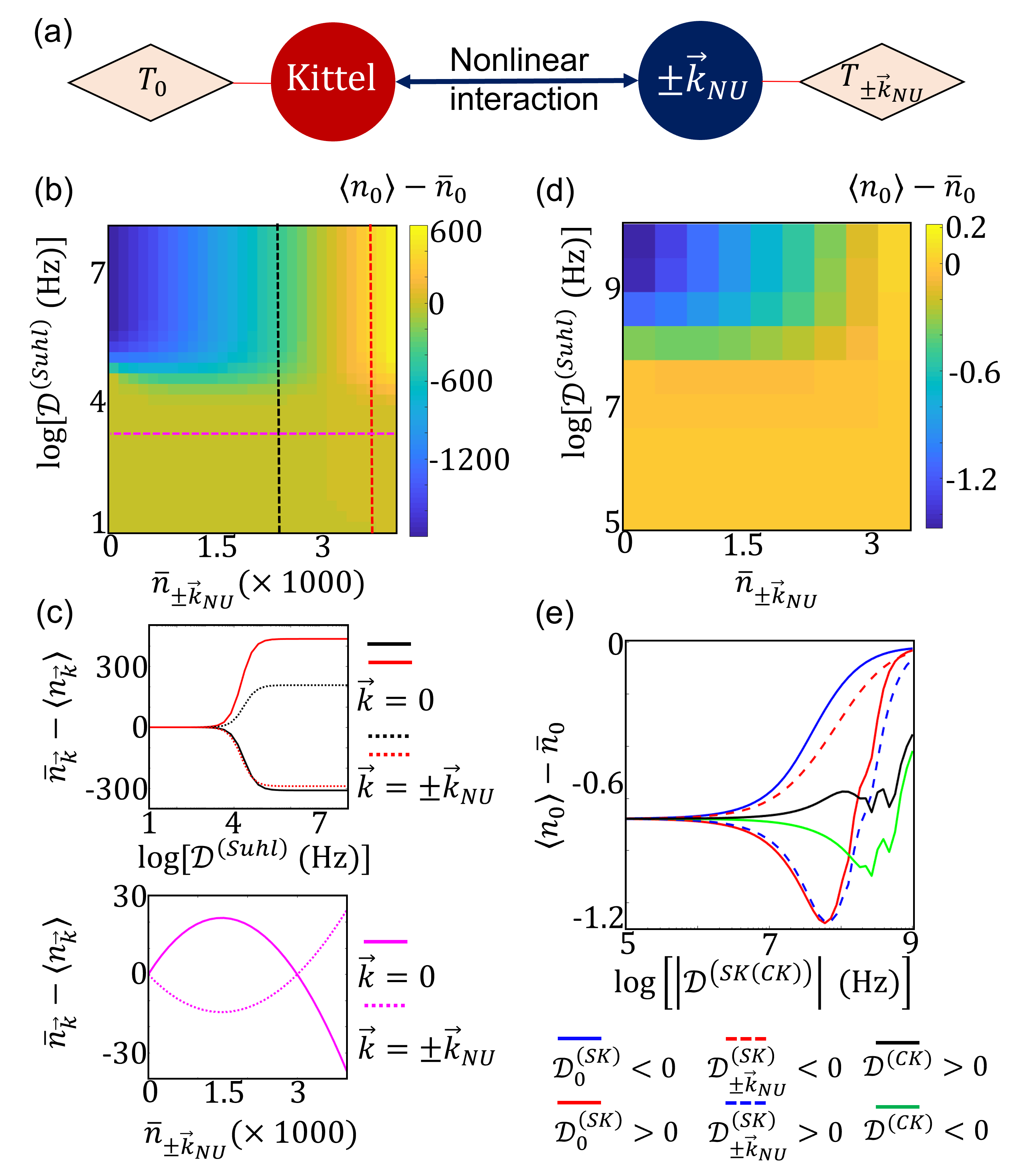}\caption{Occupation numbers
in the interacting model of Fig. \ref{fig1}(a). (a) A schematic of the Kittel
and $\pm\vec{k}_{NU}$ modes coupled to separate thermal baths. (b) $\langle
n_{0}\rangle-\bar{n}_{0}$ as a function of $\bar{n}_{\pm\vec{k}_{NU}}$ and
$|\mathcal{D}^{(\mathrm{Suhl})}|$, as calculated from Eqs. (\ref{eq15}%
)-(\ref{eq17}) for $\bar{n}_{0}=3000$. (c) Dependence of $\bar{n}_{\vec{k}%
}-\langle n_{\vec{k}}\rangle$ on selected values of $\mathcal{D}%
^{(\mathrm{Suhl})}$ (top) and $\bar{n}_{\pm\vec{k}_{NU}}$ (bottom). The black
(red) lines correspond to $\bar{n}_{\pm\vec{k}_{NU}}$ as indicated in (a). The
magenta lines correspond to $\mathcal{D}^{(\mathrm{Suhl})}$ as indicated in
(a). (d) Same as (a) but calculated from numerical solutions of $\rho^{\prime
}$ for $\bar{n}_{0}=3$. In (b)-(d), $\mathcal{D}_{\vec{k}}^{(SK)}%
=\mathcal{D}^{(CK)}=0$. (e) $\langle n_{0}\rangle-\bar{n}_{0}$ as a function
of $\mathcal{D}_{0}^{(SK)}$, $\mathcal{D}_{\pm\vec{k}_{NU}}^{(SK)}$, and
$\mathcal{D}^{(CK)}$, while $\mathcal{D}^{(\mathrm{Suhl})}=0.1\,$GHz, $\bar
{n}_{0}=3$, and $\bar{n}_{\pm\vec{k}_{NU}}=1.5$. In (b)-(e) $\xi_{0}%
/(2\pi)=\xi_{\pm\vec{k}_{NU}}/[2\pi\times(1+\alpha_{G}\Delta\omega_{\pm\vec
{k}_{NU}}/\omega_{0})]=1.5\,$MHz, $\Delta\omega_{\pm\vec{k}_{NU}}%
/(2\pi)=0.5\,$GHz.}%
\label{fig1}%
\end{figure}

For simplicity, we focus on a limited Hilbert space consisting of the Kittel
mode and a pair of magnons with opposite momenta $\pm\vec{k}_{NU}$ and
smallest detuning from the Kittel mode [see Fig. \ref{fig1}(a)], which is
motivated by the strong decay of nonlinear effects as a function of energy
differences. This model becomes better in the limit of small magnets with a
discrete spectrum. The single pair approximation is justified for the IP case
in which dipolar interaction create minima in the magnon frequency dispersion
parallel to the magnetization with the largest $|\mathcal{D}^{(\mathrm{Suhl}%
)}|$. The dispersion in the OOP is non-monotonic and isotropic, i.e. there is
no clearly dominant pair. However, the nonlinear interactions on the Kittel
mode dynamics remains the same. Since more than one pair will contribute to
the dynamics, our calculations of the RTN provide a lower bound estimate in
this case. We specify $\vec{k}_{NU}$ and associated model parameters for
relevant magnetization configurations in magnetic tunnel junctions in Section
\ref{RTN}. The reduced Hamiltonian $\mathcal{H}^{\prime}$ contains the Suhl
interaction $\mathcal{H}^{\prime(\mathrm{Suhl})}=\mathcal{D}^{(\mathrm{Suhl}%
)}c_{0}^{\dag}c_{0}^{\dag}c_{\vec{k}_{NU}}c_{-\vec{k}_{NU}}+\mathrm{H.c}.$,
the cross-Kerr interaction as $\mathcal{H}^{\prime(CK)}=\mathcal{D}%
^{(CK)}c_{0}^{\dag}c_{0}(c_{\vec{k}_{NU}}^{\dag}c_{\vec{k}_{NU}}+c_{-\vec
{k}_{NU}}^{\dag}c_{-\vec{k}_{NU}})$, and self-Kerr interaction as
$\mathcal{H}^{\prime(SK)}=\sum_{\vec{k}\in\left\{  {0,\pm\vec{k}_{NU}%
}\right\}  }\mathcal{D}_{\vec{k}}^{(SK)}c_{\vec{k}}^{\dag}c_{\vec{k}}%
c_{\vec{k}}^{\dag}c_{\vec{k}}$. We refer to $(c_{0}^{\dag}c_{0}^{\dag}%
c_{\vec{k}_{NU}}c_{-\vec{k}_{NU}}+\mathrm{H.c}.)$ as `Suhl' interaction,
because it causes the second order Suhl instability of the Kittel mode
\cite{Suhl1957,Lvov1994,Rezende2020,Kurebayashi2011,Elyasi2020,Elyasi2022,Lee2023,Sheng2023}%
. The associated density matrix is $\rho^{\prime}$. 

\subsection{\label{analytic} Analytical analysis}

A convenient measure of the density matrix $\rho^{\prime}$ is the Wigner
distribution function
\begin{align}
&  W^{\prime}(\alpha_{0},\alpha_{\vec{k}_{NU}},\alpha_{-\vec{k}_{NU}}%
)=\frac{1}{\pi^{2}}\int d^{2}\mathfrak{z}_{0}\int d^{2}\mathfrak{z}_{\vec
{k}_{NU}}\int d^{2}\mathfrak{z}_{-\vec{k}_{NU}}\nonumber\\
&  \mathrm{Tr}\left(  \rho^{\prime}\prod_{\vec{k}\in\{0,\pm\vec{k}_{NU}%
\}}e^{i\mathfrak{z}_{\vec{k}}^{\ast}c_{\vec{k}}^{\dag}}e^{i\mathfrak{z}%
_{\vec{k}}c_{\vec{k}}}\right)  \prod_{\vec{k}\in\{0,\pm\vec{k}_{NU}%
\}}e^{-i\mathfrak{z}_{\vec{k}}^{\ast}\alpha_{\vec{k}}^{\ast}}e^{-i\mathfrak{z}%
_{\vec{k}}\alpha_{\vec{k}}}, \label{eq_wigner}%
\end{align}
where $\mathfrak{z}_{\vec{k}}$ is a complex variable, and $\alpha_{\vec{k}}$
is the stochastic complex variable corresponding to $c_{\vec{k}}$.
$\text{Re}[\alpha_{\vec{k}}]$ ($\text{Im}[\alpha_{\vec{k}}]$) corresponds to
the $y^{\prime}$ ($z^{\prime}$) components of the dynamic magnetization of
mode $\vec{k}$, when the equilibrium magnetization is along $\hat{x}^{\prime}%
$.\textit{ }The equation of motion of $W^{\prime}$ can be derived from the
master equation using standard approaches \cite{Carmichael1,Carmichael2}%
.\textit{ }In Ref. \cite{Elyasi2022}, we obtained the following
Fokker-Planck-like equation (FPE) of motion%
\begin{align}
&  \frac{\partial{W}^{\prime}}{\partial t}=\left[  \mathcal{W}_{L}^{\prime
}+\mathcal{W}_{D}^{\prime}+\mathcal{W}_{Suhl}^{\prime}+\mathcal{W}%
_{SK}^{\prime}+\mathcal{W}_{CK}^{\prime}\right]  W^{\prime},\label{eq7}\\
&  \mathcal{W}_{L}^{\prime}=\sum_{\vec{k}\in\{0,\pm\vec{k}_{NU}\}}\left[
i\omega_{\vec{k}}\frac{\partial}{\partial\alpha_{\vec{k}}}\alpha_{\vec{k}%
}+\mathrm{c.c}.\right]  ,\label{eq8}\\
&  \mathcal{W}_{D}^{\prime}=\sum_{\vec{k}\in\{0,\pm\vec{k}_{NU}\}}\left[
\xi_{\vec{k}}\frac{\partial}{\partial\alpha_{\vec{k}}}\alpha_{\vec{k}}%
+\xi_{\vec{k}}(\bar{n}_{\vec{k}}+\frac{1}{2})\frac{\partial^{2}}%
{\partial\alpha_{\vec{k}}\alpha_{\vec{k}}^{\ast}}+\mathrm{c.c}.\right]
,\label{eq9}\\
&  \mathcal{W}_{Suhl}^{\prime}=i\mathcal{D}^{(Suhl)}\left[  2\frac{\partial
}{\partial\alpha_{0}}\alpha_{0}^{\ast}\alpha_{\vec{k}}\alpha_{-\vec{k}}%
-\frac{\partial}{\partial\alpha_{-\vec{k}_{NU}}^{\ast}}\alpha_{0}^{\ast
2}\alpha_{\vec{k}}-\right. \nonumber\\
&  \left.  \frac{\partial}{\partial\alpha_{\vec{k}_{NU}}^{\ast}}\alpha
_{0}^{\ast2}\alpha_{-\vec{k}}-\frac{1}{4}\frac{\partial^{3}}{\partial
\alpha_{0}^{2}\partial\alpha_{-\vec{k}_{NU}}^{\ast}}\alpha_{\vec{k}_{NU}%
}-\right. \nonumber\\
&  \left.  \frac{1}{4}\frac{\partial^{3}}{\partial\alpha_{0}^{2}\partial
\alpha_{\vec{k}_{NU}}^{\ast}}\alpha_{-\vec{k}_{NU}}+\frac{1}{2}\frac
{\partial^{3}}{\partial\alpha_{0}\partial\alpha_{\vec{k}_{NU}}^{\ast}%
\partial\alpha_{-\vec{k}_{NU}}^{\ast}}\alpha_{0}^{\ast}\right]  +\mathrm{c.c}%
.,\label{eq10}\\
&  \mathcal{W}_{SK}^{\prime}=\sum_{\vec{k}\in\{0,\pm\vec{k}_{NU}%
\}}i\mathcal{D}_{\vec{k}}^{(SK)}\left[  2\frac{\partial}{\partial\alpha
_{\vec{k}}}|\alpha_{\vec{k}}|^{2}\alpha_{\vec{k}}+\right. \nonumber\\
&  \left.  \frac{1}{2}\frac{\partial^{3}}{\partial\alpha_{\vec{k}}%
\partial\alpha_{\vec{k}}^{\ast2}}\alpha_{\vec{k}}^{\ast}\right]
+\mathrm{c.c}.,\label{eq11}\\
&  \mathcal{W}_{CK}^{\prime}=\sum_{\vec{k}\in\{\pm\vec{k}_{NU}\}}%
i\mathcal{D}^{(CK)}\left[  \frac{\partial}{\partial\alpha_{0}}|\alpha_{\vec
{k}}|^{2}\alpha_{0}+\frac{\partial}{\partial\alpha_{\vec{k}}}|\alpha_{0}%
|^{2}\alpha_{\vec{k}}+\right. \nonumber\\
&  \left.  \frac{1}{4}\frac{\partial^{2}}{\partial\alpha_{0}\alpha_{0}^{\ast
}\alpha_{\vec{k}}^{\ast}}\alpha_{\vec{k}}^{\ast}+\frac{1}{4}\frac{\partial
^{2}}{\partial\alpha_{\vec{k}}\alpha_{\vec{k}}^{\ast}\alpha_{0}^{\ast}}%
\alpha_{0}^{\ast}\right]  +\mathrm{c.c}., \label{eq12}%
\end{align}
that follows from the master equation for $\rho^{\prime}$.

In the absence of non-linearities $\mathcal{D}^{(\mathrm{Suhl})}%
=\mathcal{D}_{\vec{k}}^{(SK)}=\mathcal{D}^{(CK)}=0$, the steady state solution
of the FPE, Eq. (\ref{eq7}), becomes
\begin{equation}
W_{\mathrm{ss}}^{\prime}=\mathcal{N}\exp\left(  -\frac{2|\alpha_{0}|^{2}}%
{\bar{n}_{0}}-\frac{2|\alpha_{\vec{k}_{NU}}|^{2}}{\bar{n}_{\vec{k}_{NU}}%
}-\frac{2|\alpha_{-\vec{k}_{NU}}|^{2}}{\bar{n}_{-\vec{k}_{NU}}}\right)  ,
\label{eq13}%
\end{equation}
where $\mathcal{N}$ is a normalization constant. When $\mathcal{D}%
^{(\mathrm{Suhl})}=0$, but $\mathcal{D}_{\vec{k}}^{(SK)}\neq0$ or
$\mathcal{D}^{(CK)}\neq0$, $\mathcal{W}_{SK}^{\prime}W_{\mathrm{ss}}^{\prime
}=\mathcal{W}_{CK}^{\prime}W_{\mathrm{ss}}^{\prime}=0$, i.e., $W_{\mathrm{ss}%
}^{\prime}$ is still the steady state solution. When $\mathcal{D}%
^{(\mathrm{Suhl})}\neq0$,
\begin{align}
\mathcal{W}_{\mathrm{Suhl}}^{\prime}W_{\mathrm{ss}}^{\prime}  &
=i\mathcal{D}^{(\mathrm{Suhl})}\left(  -\frac{4}{\bar{n}_{0}}+\frac{4}{\bar
{n}_{\pm\vec{k}_{NU}}}+\frac{4}{\bar{n}_{\pm\vec{k}_{NU}}\bar{n}_{0}^{2}%
}-\right. \nonumber\\
&  \left.  \frac{4}{\bar{n}_{\pm\vec{k}_{NU}}^{2}\bar{n}_{0}}\right)
+\mathrm{c.c}. \label{eq14}%
\end{align}
According to equation (\ref{eq14}) and $\bar{n}_{0}=\bar{n}_{\pm\vec{k}_{NU}}%
$, $\mathcal{W}_{Suhl}^{\prime}W_{\mathrm{ss}}^{\prime}=0$, and
$W_{\mathrm{ss}}^{\prime}$ still holds for the steady state. However, when
$\bar{n}_{0}\neq\bar{n}_{\pm\vec{k}_{NU}}$, $\mathcal{W}_{\mathrm{Suhl}%
}^{\prime}W_{\mathrm{ss}}^{\prime}\neq0$, and $W_{\mathrm{ss}}^{\prime}$ does
not solve the problem anymore.

Since finding the steady state solution $W^{\prime}$ when $\mathcal{D}%
^{(\mathrm{Suhl})}\neq0$ is a formidable task, we focus on the magnon numbers
$\langle n_{\vec{k}}\rangle=\langle c_{\vec{k}}^{\dag}c_{\vec{k}}\rangle$. The
master equation for $\rho^{\prime}$ reduces then to
\begin{align}
&  \frac{d\mathcal{X}}{dt}=i2\Delta\omega_{\pm\vec{k}_{NU}}\mathcal{X}%
-i\mathcal{D}^{(\mathrm{Suhl})}\left[  (4\langle n_{0}\rangle+2)\langle
n_{\vec{k}_{NU}}\rangle^{2}-\right. \nonumber\\
&  \left.  (4\langle n_{\vec{k}_{NU}}\rangle+2)\langle n_{0}\rangle
^{2}\right]  -2\left(  \xi_{0}+\xi_{\pm\vec{k}_{NU}}\right)  \mathcal{X}%
,\label{eq15}\\
&  \frac{d\langle n_{0}\rangle}{dt}=-2\mathcal{D}^{(\mathrm{Suhl})}%
\mathrm{Im}\left[  \mathrm{\mathcal{X}}\right]  -2\xi_{0}(\langle n_{0}%
\rangle-\bar{n}_{0})\mathrm{,}\label{eq16}\\
&  \frac{d\langle n_{\pm\vec{k}_{NU}}\rangle}{dt}=2\mathcal{D}^{(\mathrm{Suhl}%
)}\mathrm{Im}\left[  \mathrm{\mathcal{X}}\right]  -2\xi_{\pm\vec{k}_{NU}%
}(\langle n_{\pm\vec{k}_{NU}}\rangle-\bar{n}_{\pm\vec{k}_{NU}})\mathrm{,}
\label{eq17}%
\end{align}
where $\mathcal{X}=\langle c_{0}c_{0}c_{\vec{k}_{NU}}^{\dag}c_{-\vec{k}_{NU}%
}^{\dag}\rangle$.

An analytical solution for the steady state of the EOMs Eqs. (\ref{eq15}%
)-(\ref{eq17}) can be found for small deviations $x$ in $\langle n_{0}%
\rangle=\bar{n}_{0}+x$ and $y$ in $\langle n_{\pm\vec{k}_{NU}}\rangle=\bar
{n}_{\pm\vec{k}_{NU}}+y.$ To leading order
\begin{align}
&  x=\frac{\mathcal{D}^{(\mathrm{Suhl})2}(\xi_{0}+\xi_{\pm\vec{k}_{NU}})}%
{(\xi_{0}+\xi_{\pm\vec{k}_{NU}})^{2}+\Delta\omega_{\pm\vec{k}_{NU}}^{2}}%
\frac{f(\bar{n}_{0},\bar{n}_{\pm\vec{k}_{NU}})}{g(\bar{n}_{0},\bar{n}_{\pm
\vec{k}_{NU}})}\nonumber\\
&  y=\frac{-\xi_{0}}{\xi_{\pm\vec{k}_{NU}}}x, \label{eq18}%
\end{align}
where
\begin{align}
&  f(\bar{n}_{0},\bar{n}_{\pm\vec{k}_{NU}})=2(\bar{n}_{\pm\vec{k}_{NU}}%
^{2}-\bar{n}_{0}^{2})+\nonumber\\
&  4(\bar{n}_{0}\bar{n}_{\pm\vec{k}_{NU}}^{2}-\bar{n}_{0}^{2}\bar{n}_{\pm
\vec{k}_{NU}})\nonumber\\
&  g(\bar{n}_{0},\bar{n}_{\pm\vec{k}_{NU}})=2\xi_{0}+\frac{\left(
\mathcal{D}^{(\mathrm{Suhl})}\right)  ^{2}(\xi_{0}+\xi_{\pm\vec{k}_{NU}}%
)}{(\xi_{0}+\xi_{\pm\vec{k}_{NU}})^{2}+\Delta\omega_{\pm\vec{k}_{NU}}^{2}%
}\times\nonumber\\
&  \left[  \left(  4\bar{n}_{\pm\vec{k}_{NU}}^{2}-8\bar{n}_{0}\bar{n}_{\pm
\vec{k}_{NU}}-4\bar{n}_{0}\right)  +\frac{\xi_{0}}{\xi_{\pm\vec{k}_{NU}}%
}\times\right. \nonumber\\
&  \left.  \left(  4\bar{n}_{0}^{2}-8\bar{n}_{0}\bar{n}_{\pm\vec{k}_{NU}%
}-4\bar{n}_{\pm\vec{k}_{NU}}\right)  \right]  . \label{eq19}%
\end{align}
When $\bar{n}_{0}=\bar{n}_{\pm\vec{k}_{NU}}=0$, $f(\bar{n}_{0},\bar{n}%
_{\pm\vec{k}_{NU}})=0$, thus $x=y=0$ as expected. When $\bar{n}_{0}>\bar
{n}_{\pm\vec{k}_{NU}}$ ($\bar{n}_{0}<\bar{n}_{\pm\vec{k}_{NU}}$), $x<0$
($x>0$) and $y>0$ ($y<0$), i.e., the Kittel and the $\pm\vec{k}_{NU}$ magnons
equilibrate towards temperatures that are governed by the reservoirs.

\subsection{\label{results} Numerical results and discussion}

In the steady state, Eqs. (\ref{eq15})-(\ref{eq17}) reduce to a polynomial
equation in $\langle n_{\pm\vec{k}_{NU}}\rangle$ and $\langle n_{0}\rangle$.
Figure \ref{fig1}(a) shows the dependence of $\langle n_{0}\rangle-\bar{n}%
_{0}$ for fixed $\bar{n}_{0}$ as a function of $\bar{n}_{\pm\vec{k}_{NU}}$ and
$\mathcal{D}^{(\mathrm{Suhl})}$ ($\mathcal{D}_{\vec{k}}^{(SK)}=\mathcal{D}%
^{(CK)}=0)$. When $\bar{n}_{0}>\bar{n}_{\pm\vec{k}_{NU}}$ ($\bar{n}_{0}%
<\bar{n}_{\pm\vec{k}_{NU}}$) we find that $\langle n_{0}\rangle<\bar{n}_{0}$
($\langle n_{0}\rangle>\bar{n}_{0}$), as expected from the analysis in Sec.
\ref{analytic} and Eqs. (\ref{eq18})-(\ref{eq19}). For fixed $\bar{n}_{\pm
\vec{k}_{NU}}$, $|\langle n_{0}\rangle-\bar{n}_{0}|$ increases and eventually
saturates with increasing $|\mathcal{D}^{(\mathrm{Suhl})}|$, approaching
complete equilibration between the Kittel mode and the magnon pair to\ a
common temperature. Figure \ref{fig1}(c) shows examples of $\langle n_{\vec
{k}}\rangle-\bar{n}_{\vec{k}}$ for both $\vec{k}=0$ and $\vec{k}=\pm\vec
{k}_{NU}$ as functions of $\mathcal{D}^{(\mathrm{Suhl})}$ and $\bar{n}%
_{\pm\vec{k}_{NU}}$, at $\bar{n}_{\pm\vec{k}_{NU}}$ and $\mathcal{D}%
^{(\mathrm{Suhl})}$ values as indicated by the straight lines of the same
color in Fig. \ref{fig1}(b).

We must keep in mind that approximations are valid only for temperatures that
correspond to small $\bar{n}_{0}$ and $\bar{n}_{\pm\vec{k}_{NU}}$. Figure
\ref{fig1}(c) and Fig. \ref{fig1}(a) are similar (except for $\bar{n}_{0}=3)$,
as expected. Figure \ref{fig1}(e) shows $\langle n_{0}\rangle-\bar{n}_{0}$ as
a function of $\mathcal{D}_{0}^{(SK)}$, $\mathcal{D}_{\pm\vec{k}_{NU}}^{(SK)}%
$, and $\mathcal{D}^{(CK)}$ for a fixed $\mathcal{D}^{(\mathrm{Suhl})}\neq0$,
and $\bar{n}_{0}\neq\bar{n}_{\pm\vec{k}_{NU}}$. The self-Kerr and cross-Kerr
interactions lead to the frequency shifts $\Delta\omega_{\pm\vec{k}_{NU}}$.An
increase (decrease) in $\Delta\omega_{\pm\vec{k}_{NU}}$ naturally decreases
(increases) $|\langle n_{0}\rangle-\bar{n}_{0}|$ [see Eq. (\ref{eq18})].

\begin{figure}[ptb]
\includegraphics[width=0.5\textwidth]{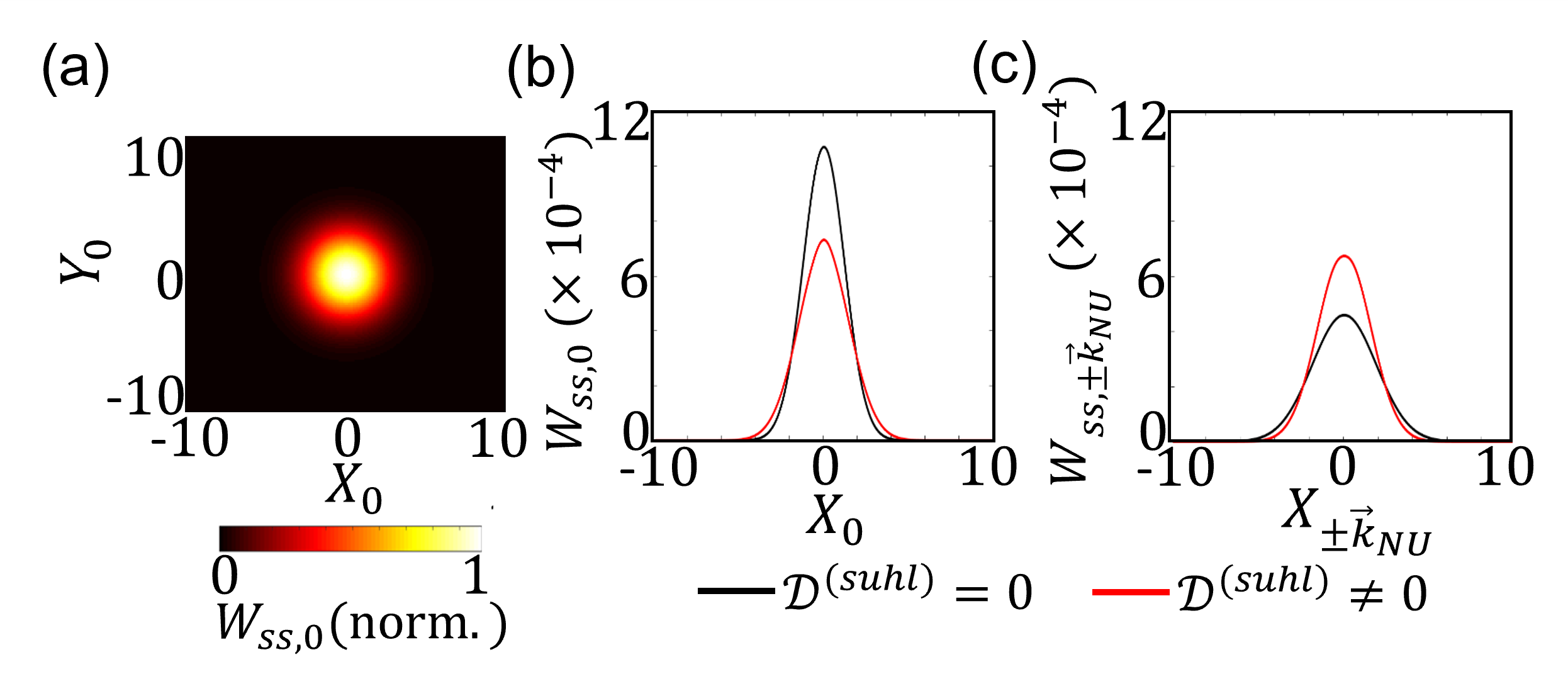}\caption{Wigner function of
three-level magnon system. (a) $W_{ss,0}^{\prime}(X_{0},Y_{0})$, when $\bar
{n}_{0}=0$, and $\mathcal{D}^{(\mathrm{Suhl})}=0$. (b) and (c) $W_{ss,0}%
^{\prime}$ and $W_{\pm\vec{k}_{NU}}^{\prime}$, respectively, for
$\mathcal{D}^{(Suhl)}=0$ and $\mathcal{D}^{(\mathrm{Suhl})}=0.1\,$GHz,
$\bar{n}_{0}=3$, $\bar{n}_{\pm\vec{k}_{NU}}=1.5$, $\mathcal{D}^{(\mathrm{Suhl}%
)}=0.1\,$GHz, and $\mathcal{D}_{\vec{k}}^{(SK)}=\mathcal{D}^{(CK)}=0$. The
other parameters are those used in Fig. \ref{fig1}.}%
\label{fig2}%
\end{figure}

Figure \ref{fig2}(a) shows the equilibrium Wigner function of the Kittel mode,
$W^{\prime}_{ss,0}(X_{0},Y_{0})$ for $\mathcal{D}^{(\mathrm{Suhl})}=0$, where
$W^{\prime}_{ss,\vec{k}}(X_{\vec{k}},Y_{\vec{k}})$ is the steady state Wigner
function of $\vec{k}$ mode after tracing out the other modes, $X_{\vec{k}%
}=\mathrm{Re}[\alpha_{\vec{k}}]$ and $Y_{\vec{k}}=\mathrm{Im}[\alpha_{\vec{k}%
}]$. The latter variables represent the dynamic magnetizations of modes with
index $\vec{k}$ by $\vec{M}_{\vec{k}}\cdot\hat{y}^{\prime}=M_{s}X_{\vec{k}%
}/\sqrt{S}$ and $\vec{M}_{\vec{k}}\cdot\hat{z}^{\prime}=M_{s}Y_{\vec{k}}%
/\sqrt{S}$ in a rotating frame, where $S$ is the total spin and $M_{s}$ is the
saturation magnetization. Figures \ref{fig2}(b) and (c) show $W_{ss,0}%
(X_{0},0)$ and $W_{ss,\pm\vec{k}_{NU}}(X_{\pm\vec{k}_{NU}},0)$ (steady state
Wigner function of either of the $\pm\vec{k}_{NU}$ modes), respectively, for
$\mathcal{D}^{(\mathrm{Suhl})}=0$ (black lines) and $\mathcal{D}%
^{(\mathrm{Suhl})}\neq0$ (red lines). Since $\bar{n}_{0}<\bar{n}_{\pm\vec
{k}_{NU}}$, the magnon numbers equilibrate by shrinking (expanding) the Wigner
function (relative to is form at $\mathcal{D}^{(\mathrm{Suhl})}=0)$ of the
$\vec{k}_{NU}$ (Kittel) mode. In the following, we show how the distributions
functions affect the RTN and switching frequency.

\section{\label{RTN} Random telegraph noise by magnetization reversal}

\subsection{\label{past} Introduction}

\subsubsection{Magnetic tunnel junctions and magnon parametron}

In Sec. \ref{eqb_dist}, we addressed the FPE-like EOM for the Wigner function
[see Eq. (\ref{eq7})] of interacting magnons. Without a drive, the effective
potential of each of the magnon modes has only one minimum. While not directly
relevant for the present generation of experiments on MTJ, we point out here
the relation with the \textquotedblleft magnon parametron\textquotedblright, a
magnetic disk that becomes bistable under a parametric excitation of the
Kittel mode \cite{Makiuchi2021,Elyasi2022}. Its two potential minima can be
tuned into the stochastic switching regime \cite{Makiuchi2021,Elyasi2022}. The
distribution function $W_{\vec{m}}$ of the magnetization $\vec{m}$ with
constant modulus around each of the two minima is similar to the equilibrium
$W_{ss,0}^{\prime}$ (Wigner function of the Kittel mode steady state) derived
above, but entails additional dynamical effects as demonstrated in this
section. The FPE EOM for the distribution function $W_{\vec{m}}$ has the form
\begin{equation}
\frac{\partial W_{\vec{m}}}{\partial t}=\left[  -\frac{\partial}{\partial
x_{i}}\mathcal{A}_{i}(\vec{x})+\frac{1}{2}\frac{\partial^{2}}{\partial
x_{i}\partial x_{j}}\mathcal{B}_{ij}(\vec{x})+\dots\right]  W_{\vec{m}},
\label{eq20}%
\end{equation}
where a summation over the repeated indices is implied, $\mathcal{A}_{i}$ and
$\mathcal{B}_{ij}$ are model-dependent constants that govern the drift and
diffusion terms, respectively, $\{i,j\}\in\{\theta,\phi\}$, and $\theta$
($\phi$) is the polar (azimuthal) angle of the magnetization. We can solve Eq.
(\ref{eq20}) by the ansatz
\begin{equation}
W_{\vec{m}}=W_{ss,\vec{m}}+\sum_{n}\mathcal{F}_{n}(\theta,\phi)e^{-a_{n}t},
\label{eq21}%
\end{equation}
where $W_{ss,\vec{m}}$ is the steady state solution, while the $a_{n}$ with
the largest real part, $a_{1}$, corresponds to the switching frequency
\cite{Kinsler1991,Brown1963,Elyasi2022}, i.e., $f_{s}=a_{1}$. Solving
$W_{\vec{m}}$ exactly is tedious and often impossible without additional
approximations such as Kramers escape \cite{Kramers1940,Landauer1961} or
high-barrier limit \cite{Langer1969} assumptions, which we briefly review below.

\subsubsection{Brown theory for the Kramers escape of macrospin}

The Ne\'{e}l-Brown theory considers a macrospin with free energy
$E(\theta,\phi)$, where $\theta$ is the angle with respect to the easy axis
$\hat{x}$ direction and $\phi$ is the azimuthal angle measured from the $xy$
plane. The Kramers method is valid in the high energy barrier limit in the
path of least action between the energy mimima. Here, we briefly review the
Brown theory for a $\phi$-independent free energy $E=E(\theta)$, with two
minima at $\theta_{1}=0$ and $\theta_{2}=\pi$ and a saddle point (maximum in
this 1D case) at $0<\theta_{sd}<\pi$. The critical assumption is a Maxwell
distribution for $[0\,\theta_{r1}]([\theta_{r2}\,\pi])$, where $0(\theta
_{sd})<\theta_{r1}(\theta_{r2})<\theta_{sd}(\pi)$, $W\approx W_{1(2)}%
=W_{1(2)}(0(\pi))\exp\{-\beta\lbrack E_{1(2)}(\theta)-E(0(\pi))]\}$,
$\beta=V_{s}/k_{B}T$, $V_{s}$ is the sample volume, $E_{1(2)}(\theta)\approx
E(0(\pi))+E^{\prime\prime}(0(\pi))\theta^{2}$, and $E^{\prime\prime}%
(\theta)=(\partial^{2}E/\partial\theta^{2})|_{\theta}$. The total number of
magnons close to the minima $[0\,\theta_{r1}]([\theta_{r2}\,\pi])$ is
$N_{1(2)}=\int_{0(\theta_{2})}^{\theta_{1}(\pi)}W_{1(2)}d\theta$. The integral
can be carried out under the high barrier assumption or $\theta_{r1}%
\rightarrow\infty$ $(\theta_{r2}\rightarrow-\infty)$ and $N_{1(2)}\approx
W_{1(2)}(0(\pi))\times\lbrack\beta E^{\prime\prime}(0(\pi))]^{-1}$.

The probability current $\vec{J}$ through the saddle point is conserved and
related to the distribution function by $\partial W/\partial t=-\nabla
\cdot\vec{J}$ that in our case leads to $J_{\theta}=-\mathcal{A}_{\theta
}W+\frac{1}{2}\mathcal{B}_{\theta,\theta}\partial W/\partial\theta$. Brown
derived that $\mathcal{A}_{\theta}=-\mathfrak{a}\partial E/\partial\theta$ and
$\mathcal{B}_{\theta,\theta}=2\mathfrak{b}=\mathfrak{a}/\beta$, and
$\mathfrak{a}=\alpha_{G}\times\lbrack\gamma^{-2}+(\alpha_{G}M_{s})^{2}]^{-1}$,
$\gamma$ is the gyromagnetic constant and $M_{s}$ is the saturation
magnetization. After multiplying the latter with $\exp[\beta E(\theta)]$ and
integrating over $\theta$, $J_{\theta}=-\dot{N}_{1}=\dot{N}_{2}=-v_{21}%
N_{2}+v_{12}N_{1}$, where
\begin{equation}
v_{12}(v_{21})=\mathfrak{a}\sqrt{\beta\lambda_{sd}/(2\pi)}\lambda_{0(\pi)}%
\exp\{-\beta\lbrack E(\theta_{sd})-E(0(\pi))]\},\label{eq22}%
\end{equation}
and $\lambda_{x}=E^{\prime\prime}(x)$.

This approach can be extended to a $\phi$-dependent system as long as it
supports two (meta)stable states at $\theta_{1}$ and $\theta_{2}$ and a saddle
node $\theta_{sd}$ that we assume to be at $\phi=0$ without loss of
generality, thereby reproducing \cite{Brown1979}
\begin{equation}
v_{ij}=\mathcal{G}\frac{\alpha_{G}\gamma}{M_{s}(1+\alpha_{G}^{2})}\sqrt
{\frac{-\lambda_{sd,\phi}}{\lambda_{sd,\theta}}}\sqrt{\lambda_{i,\phi}%
\lambda_{i,\theta}}e^{-\beta\lbrack E(\theta_{sd})-E(\theta_{i})]},
\label{eq23}%
\end{equation}
where $\mathcal{G}=(-2\lambda_{sd,\phi})^{-1}\times\{(-\lambda_{sd,\phi
}-\lambda_{sd,\theta})+[(-\lambda_{sd,\phi}-\lambda_{sd,\theta})^{2}%
-4\lambda_{sd,\phi}\lambda_{sd,\theta}/\alpha_{G}^{2}]^{1/2}\}$,
$\lambda_{x,y}=(\partial^{2}E/\partial y^{2})|_{\theta_{x}}$. The switching
frequency is therefore
\begin{equation}
f_{s}=a_{1}=v_{12}+v_{21}. \label{eq24}%
\end{equation}

\subsubsection{Braun theory for nonuniform magnetization switching}

The N\'{e}el-Brown theory relies on a magnetization that remains spatially uniform
during the switching process. However, depending on the shape and size of the
magnet, the applied magnetic field, and material parameters such as
crystalline anisotropy and exchange length, the saddle point in the free
energy may belong to a magnetic texture, even when the magnetization of the
minima are uniform. For a wire with magnetic fields applied along the easy
axis that are small compared with the anisotropy, Braun
\cite{Braun1993,Braun1994} found saddle points of the free energy for two
domain walls and derived an FPE starting from the linearized Landau-Lifshitz-Gilbert (LLG) with thermal
fluctuations. The stochastic switching rate follows then from an extension of
the Langer's theory for metastable decay \cite{Langer1969} in the high barrier
limit and assuming an equilibrium distributions $W_{2}=0$ around the stable
state and $W_{1}=W_{eqb}=Z^{-1}e^{-\beta E_{sd}^{(2)}}$ around the metastable
one, where $E_{sd}^{(2)}$ is the expansion of free energy at the saddle point
up to second order in $\theta$ $(\phi)$ of the fluctuations $p$ $(q)$. This
approximation assumes that all magnons quasiparticles reside in the metastable
region while none exist beyond the saddle point, i.e., $\int W_{eqb}dpdq=1$
which fixes the partition function $Z$. The nonequilibrium distribution
becomes $W=FW_{eqb}$, and $F$ can be derived from the FPE, from which the
probability current $\vec{J}$ follows. The switching frequency is obtained by
integration of $\vec{J}$ over over all the dimensions transverse to the path
of least action. For a cylindrical\textit{ }wire \cite{Braun1994}
\begin{equation}
f_{s}=\mathcal{P}\sqrt{\frac{\det({\mathcal{H}_{m,q})}\det({\mathcal{H}%
_{m,p})}}{\det^{\prime}(|\mathcal{H}_{sd,q}|)\det(\mathcal{H}_{sd,p})}},
\label{eq25}%
\end{equation}
where $\mathcal{H}_{sd(m),q(p)}$ are the second order expansion terms of the
free energy in $p$ and $q$ close to the extrema, i.e. $E_{sd(m)}^{q(p)}=\int
dx[p\mathcal{H}_{sd(m),p}p+q\mathcal{H}_{sd(m),q}q]$. $\det(\mathcal{H}%
_{sd(m),p(q)})=\prod_{i\in\{bs,\vec{k}\}}\lambda_{sd(m),i}^{(p(q))}$, where
$\lambda_{sd(m),i}^{p(q)}$ are the eigenvalues of $\mathcal{H}_{sd(m),p(q)}$.
`bs' stands for bound states, and prime in $\det^{\prime}$ denotes exclusion
of $\lambda_{sd(m),i}=0$, i.e. the Goldstone modes corresponding to the
translational invariance of the saddle node. $\mathcal{P}=\lambda_{d}%
\sqrt{E_{sd}-E_{m}}L\sqrt{\beta^{\prime}\mathcal{S}/2\pi^{3}}$, where
$\mathcal{S}$ ($L$) is the cross section (length) of the sample,
$\beta^{\prime}=1/(k_{B}T)$, and only $\lambda_{d}>0$ depends on the dynamics
governed by the linearized LLG at the saddle node, viz. $\lambda_{d}\chi
_{q,d}=-H_{sd,p}\chi_{p,d}-\alpha_{G}H_{sd,q}\chi_{q,d}$ and $\lambda_{d}%
\chi_{p,d}=H_{sd,q}\chi_{q,d}-\alpha_{G}H_{sd,p}\chi_{p,d}$, and $(\chi
_{p,d},\chi_{q,d})$ is the corresponding eigenfunction.

\subsubsection{Magnon interactions}

The theoretical approaches to compute $f_{s}$ as summarized above linearize
the free energy around the extrema. In Braun theory for nonuniform domain wall
assisted magnetization switching, spin waves are assumed to be
non-interacting. To the best of our knowledge, magnon interactions on the
magnetization switching and RTN have generally been neglected. In Sec.
\ref{model1}, we introduce a minimal model for the effects of magnon
interactions on the uniform magnetization dynamics. In Sec.
\ref{results1_analytic}, we discuss the phenomenology of nonlinear
interactions in the RTN. In Sec. \ref{results1_calc}, we use the model from
Sec. \ref{model1} to numerically calculate and interpret the effect of
four-magnon interactions on $f_{s}$. A full numerical solution of the LLG
equation including thermal noise may be used to test our phenomenology.

\begin{figure}[ptb]
\includegraphics[width=0.5\textwidth]{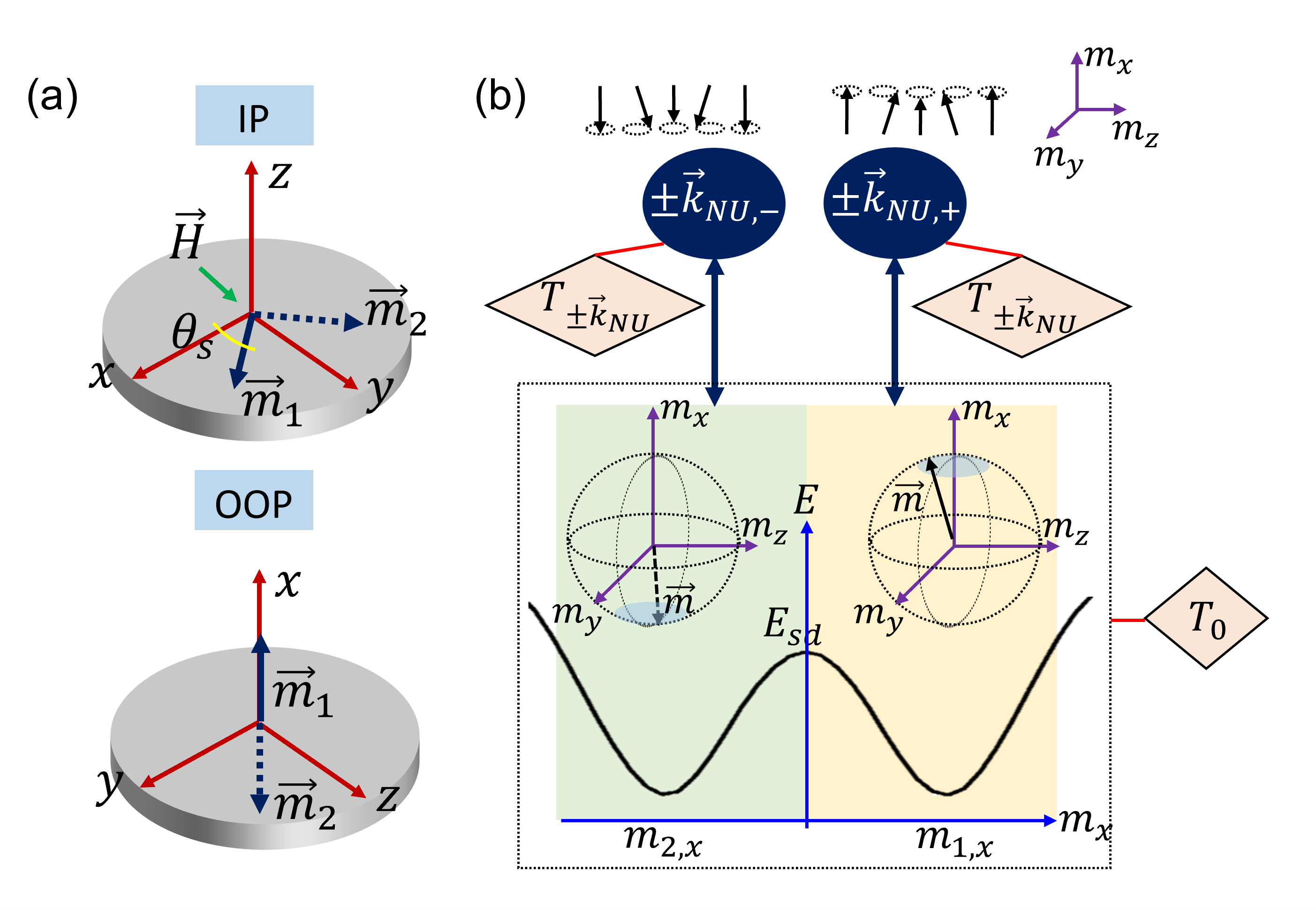}\caption{Schematics of the
two configurations of MTJs used to study stochastic switching. (a) Top panel:
in-plane configuration, with easy axis anisotropy along $\hat{x}$, hard axis
uniaxial anisotropy along $\hat{z}$, and a magnetic field applied along
$\hat{y}$. Bottom panel: the out-of-plane (OOP) configuration, with easy axis
uniaxial anisotropy along $\hat{x}$. In top and bottom panels, $\vec{m}_{1}$
and $\vec{m}_{2}$ are the equilibrium magnetization directions (minima of
macrospin free energy). (b) A sketch of the model in Sec. \ref{model1}
emphasizing the difference of the three level systems in the two minima of the
free energy. }%
\label{fig3}%
\end{figure}

\subsection{\label{model1} Model}

The stochastic switching of MTJs is explored for the in-plane (IP) and
out-of-plane (OOP) configurations as sketched in Fig. \ref{fig3}(a)
\cite{Borders2019,Hayakawa2021,Kanai2021,Kaneko2024}. We chose two coordinate
systems such that the equilibrium magnetization without an applied field is
always along the $\hat{x}$ axis. In the IP configuration, the magnetization
lies IP by a hard uniaxial anisotropy $K_{z}$ along $\hat{z}$. while an
elliptic shape defines an easy axis anisotropy $K_{x}$ along $\hat{x}$. An IP
applied magnetic field applied along $\hat{y}$ then forces a finite angle
$\theta_{s}$ between the equilibrium magnetization and $\hat{x}$ [see Fig.
\ref{fig1}(a)]. In the OOP configuration the magnetization is perpendicular to
the plane, which in our definition defined the $\hat{x}$ axis. This implies an
easy axis uniaxial anisotropy $K_{x}$ along $\hat{x}$. The free energy of the
OOP macrospin is independent of azimuthal angle, while for the IP macrospin it
depends on both the polar and azimuthal angles, which leads to different
switching characteristics for the two configurations
\cite{Brown1963,Brown1979,Kanai2021,Hayakawa2021}. Additionally, the magnon
dispersion and the relevant interaction coefficients are different for the
two. For example, for the IP case, the magnons with wavevector parallel to the
magnetization have lower frequency than the perpendicular ones, while for the
OOP case, the frequency of the magnons with in-plane wavevector does not
depend on the in-plane angle. In Ref. \cite{Kanai2023}, we already simulated
the dynamics of the uniform magnetization (macrospin) coupled to a pair of
spin waves with opposite in-plane wave vectors $\pm{\vec{k}}_{NU,i}$. Here
$i\in\{+,-\}$ indicates the configuration space around the minima with
$m_{x}>0$ and $m_{x}<0$, respectively. We pick a pair of magnons for each of
the two energy wells as sketched in Fig. \ref{fig3}(b). The interaction with
spin waves ${\vec{k}}_{NU}$ generates an effective stochastic magnetic field
$\vec{H}_{SW}=\vec{H}_{SW}^{(\mathrm{Suhl})}+\vec{H}_{SW}^{(CK)}$ on the
macrospin by the 4-magnon interactions, where
\begin{align}
H_{\mathrm{SW},x}^{(\mathrm{Suhl})} &  =\mathcal{C}\left\{  \mathrm{Re}\left[
{\alpha}_{{\vec{k}_{NU}}^{\prime}}{\alpha}_{-{\vec{k}_{NU}}^{\prime}}\right]
\left\{  \mathcal{K}{\sin{\theta}_{s}\ }\times\left[  m_{x}\sin\theta
_{s}-\right.  \right.  \right.  \label{eq26}\\
&  \left.  \left.  \left.  \mathrm{sign}\left(  m_{x}\right)  m_{y}\cos
\theta_{s}\right]  +{\mathcal{K}}^{\prime}\cos\theta_{s}\mathrm{sign}\left(
m_{x}\right)  \ \right\}  -\right.  \nonumber\\
&  \left.  \mathrm{Im}\left[  {\alpha}_{{\vec{k}_{NU}}^{\prime}}{\alpha
}_{-{\vec{k}_{NU}}^{\prime}}\right]  {\sin\theta_{s}\ }m_{z}\right\}
,\nonumber\\
&  H_{\mathrm{SW,}y}^{(\mathrm{Suhl})}=\mathcal{C}\left\{  \mathrm{Re}\left[
{\alpha}_{{\vec{k}_{NU}}^{\prime}}{\alpha}_{-{\vec{k}_{NU}}^{\prime}}\right]
\left\{  \mathcal{K}{\cos\theta_{s}\ }\times\left[  m_{y}{\cos\theta_{s}%
\ }-\right.  \right.  \right.  \nonumber\\
&  \left.  \left.  \left.  \left\vert m_{x}\right\vert \right.  \left.
{\sin\theta_{s}\ }\right]  +{\mathcal{K}}^{\prime}{\sin\theta_{s}\ }\right\}
+\mathrm{Im}\left[  {\alpha}_{{\vec{k}_{NU}}^{\prime}}{\alpha}_{-{\vec{k}%
_{NU}}^{\prime}}\right]  {\cos\theta_{s}\ }m_{z}\right\}  ,\nonumber\\
&  H_{\mathrm{SW},z}^{(\mathrm{Suhl})}=\mathcal{C}\left\{  -\mathrm{Im}\left[
{\alpha}_{{\vec{k}_{NU}}^{\prime}}{\alpha}_{-{\vec{k}_{NU}}^{\prime}}\right]
\left[  m_{x}{\sin\theta_{s}\ }-\mathrm{sign}\left(  m_{x}\right)  -\right.
\right.  \nonumber\\
&  \left.  \left.  m_{y}{\cos\theta_{s}\ }\right]  \mathcal{K}\times
\mathrm{Re}\left[  {\alpha}_{{\vec{k}_{NU}}^{\prime}}{\alpha}_{-{\vec{k}_{NU}%
}^{\prime}}\right]  m_{z}\right\}  ,\label{eq27}\\
&  H_{\mathrm{SW},x}^{(CK)}={\mathcal{C}}^{\prime}{\left\vert {\alpha}%
_{\pm{\vec{k}_{NU}}^{\prime}}\right\vert }^{2}\{-\mathcal{K}\times
\mathrm{sign}\left(  m_{x}\right)  {\cos\theta_{s}\ }-2{\mathcal{K}}%
^{\prime\prime}\times\nonumber\\
&  [2{{\sin}^{2}\theta_{s}\ }m_{x}-2{\sin\theta_{s}\ }{\cos\theta_{s}%
\ }\mathrm{sign}\left(  m_{x}\right)  m_{y}]\},\nonumber\\
&  H_{\mathrm{SW},y}^{(CK)}={\mathcal{C}}^{\prime}{\left\vert {\alpha}%
_{\pm{\vec{k}_{NU}}^{\prime}}\right\vert }^{2}\{-\mathcal{K}\times{\sin
\theta_{s}\ }-2{\mathcal{K}}^{\prime\prime}\times\nonumber\\
&  [2{{\cos}^{2}{\theta}_{s}\ }m_{y}-2{\sin\theta_{s}\ }{\cos\theta_{s}%
\ }\left\vert m_{x}\right\vert ]\},\nonumber\\
&  H_{\mathrm{SW,}z}^{(CK)}={\mathcal{C}}^{\prime}{\left\vert {\alpha}%
_{\pm{\vec{k}_{NU}}^{\prime}}\right\vert }^{2}\{-2{\mathcal{K}}^{\prime\prime
}\times\lbrack-2m_{z}]\},\label{eq28}%
\end{align}
and $\vec{H}_{SW}^{(Suhl)}$ and $\vec{H}_{SW}^{(CK)}$ derive from the
Hamiltonians $\mathcal{H}^{\prime(Suhl)}$ and $\mathcal{H}^{\prime(CK)}$,
respectively. Here $\mathcal{C}={-4{\mathcal{D}}_{\pm\vec{k}_{NU}}^{(Suhl)}%
}/{\gamma\left(  1+\left\vert m_{x}\right\vert \cos\theta_{s}+m_{y}\sin
\theta_{s}\right)  }$, $\mathcal{K}=(u_{0}^{2}+v_{0}^{2})$, and ${\mathcal{K}%
}^{\prime}=u_{0}v_{0}\left(  1+\left\vert m_{x}\right\vert \cos\theta
_{s}+m_{y}\sin\theta_{s}\right)  $. ${\vec{k}_{NU}^{\prime}}={\vec{k}}_{NU,+}$
(${\vec{k}_{NU}^{\prime}}={\vec{k}}_{NU,-}$) when $m_{x}>0$ ($m_{x}<0$).
${\mathcal{C}}^{\prime}=2{\mathcal{D}}^{(CK)}/\gamma$ and ${\mathcal{K}%
}^{\prime\prime}=u_{0}v_{0}/\left(  1+\left\vert m_{x}\right\vert \cos
\theta_{s}+m_{y}\sin\theta_{s}\right)  $. $u_{0}=\sqrt{(A_{0}+\omega
_{0})/(2{\omega}_{0})},$ $v_{0}=-\mathrm{sign}(B_{0})\sqrt{(A_{0}-{\omega}%
_{0})/(2{\omega}_{0})}$, $A_{0}=|\gamma{\mu}_{0}H_{y}|+{\omega}_{M}(K_{x}%
\sin^{2}\theta_{s}+K_{z})/2$, $B_{0}=\omega_{M}(K_{x}{\sin}^{2}\theta
_{s}-K_{z})/2$, ${\omega}_{0}=\sqrt{A_{0}^{2}-|B_{0}{|}^{2}}$, and $\omega
_{M}=\gamma M_{s}$.$S=V_{s}M_{s}/\left(  h\gamma\right)  $ is the total number
of spins in the magnet.

Vice versa, the dynamics of the macrospin parametrically excites the spin wave
pairs via the Hamiltonian $\mathcal{H}^{\prime(\mathrm{Suhl})}\equiv\sum
_{i}P_{i}c_{{\vec{k}}_{NU,i}}^{\dagger}c_{-{\vec{k}}_{NU,i}}^{\dagger
}+\mathrm{H.c}.$, where
\begin{align}
P_{i}  &  =-\frac{S\gamma{\mathcal{C}}_{i}}{2}\left\{  \mathcal{K}%
\times\left[  {\left(  \left\vert m_{x}\right\vert {\sin\theta_{s}}-m_{y}%
{\cos\theta_{s}}\right)  }^{2}\right.  \left.  -m_{z}^{2}\right]  +\right.
\nonumber\\
&  \left.  2i\left[  -m_{x}m_{z}{\sin\theta_{s}}+m_{y}m_{z}\mathrm{sign}%
\left(  m_{x}\right)  {\cos\theta_{s}}\right]  -\right. \nonumber\\
&  \left.  2{\mathcal{K}}^{\prime}\left[  1-\left\vert m_{x}\right\vert
\cos\theta_{s}-m_{y}\sin\theta_{s}\right]  \right\}  , \label{eq29}%
\end{align}
${\mathcal{C}}_{+(-)}=\mathcal{C}$ when $m_{x}>0$ ($m_{x}<0$) and zero
otherwise. The cross-Kerr interaction $\mathcal{H}^{\prime(CK)}\equiv
\sum_{\vec{k}\in\{0,\pm\vec{k}_{NU,+(-)}\}}\omega_{\vec{k}}^{\prime}c_{\vec
{k}}^{\dag}c_{\vec{k}}$ leads to the dynamical frequency shift of the spin
waves
\begin{align}
&  \Delta{\omega^{\prime}}_{\pm{\vec{k}}_{NU,i}}={\mathcal{D}}_{i}%
^{(CK)}\times\left\{  S\mathcal{K}\left(  1-\right.  \right. \nonumber\\
&  \left.  \left.  \cos\theta_{s}\left\vert m_{x}\right\vert -{\sin\theta_{s}%
}m_{y}\right)  -2S^{2}{\mathcal{K}}^{\prime}\left[  {{\sin}^{2}\theta_{s}%
}m_{x}^{2}+\right.  \right. \nonumber\\
&  \left.  \left.  {{\cos}^{2}\theta_{s}}m_{y}^{2}-2{\sin\theta_{s}}%
{\cos\theta_{s}}\left\vert m_{x}\right\vert m_{y}-m_{z}^{2}\right]  \right\}
, \label{eq30}%
\end{align}
where ${\mathcal{D}}_{+(-)}^{(CK)}={\mathcal{D}}^{(CK)}$ when $m_{x}>0$
($m_{x}<0$) otherwise zero. We can now formulate the coupled equations of
motion that consist of the stochastic LLG equations
for the Kittel mode%
\begin{align}
&  \frac{d\vec{m}}{dt}=-\gamma{\mu}_{0}\vec{m}\times\left(  {\vec{H}%
}_{\mathrm{A}}+{\vec{H}}_{\mathrm{dc}}+{\vec{H}}_{\mathrm{SW}}+{\vec{H}%
}_{\mathrm{th}}\right)  +\\
&  {\alpha}_{\mathrm{G}}\vec{m}\times\frac{d\vec{m}}{dt},\nonumber
\end{align}
while following $\dot{W}^{\prime}$ (see Sec. \ref{analytic}), the spin wave
pair obeys
\begin{align}
&  \frac{dX_{\pm{\vec{k}}_{NU,i}}}{dt}=-({\omega}_{\pm{\vec{k}}_{NU,i}%
}^{\prime}+{\omega}_{\pm{\vec{k}}_{NU,i}})Y_{\pm{\vec{k}}_{NU,i}}-\\
&  {\alpha}_{\mathrm{G}}{\omega}_{\pm{\vec{k}}_{NU,i}}X_{\pm{\vec{k}}_{NU,i}%
}+F_{th}+\nonumber\\
&  \frac{i{\mathcal{D}}_{\pm{\vec{k}}_{NU,i}}^{(Suhl)}}{2}\left(  P_{i}%
{\alpha}_{\mp{\vec{k}}_{NU,i}}^{\ast}-P_{i}^{\ast}{\alpha}_{\mp{\vec{k}%
}_{NU,i}}\right)  +\nonumber\\
&  i{\mathcal{D}}_{\pm{\vec{k}}_{NU,i}}^{(SK)}\left(  {\alpha}_{\pm{\vec{k}%
}_{NU,i}}^{\ast}{\alpha}_{\pm{\vec{k}}_{NU,i}}{\alpha}_{\pm{\vec{k}}_{NU,i}%
}-\right. \nonumber\\
&  \left.  {\alpha}_{\pm{\vec{k}}_{NU,i}}{\alpha}_{\pm{\vec{k}}_{NU,i}}^{\ast
}{\alpha}_{\pm{\vec{k}}_{NU,i}}^{\ast}\right)  ,\nonumber
\end{align}%
\begin{align}
&  \frac{dY_{\pm{\vec{k}}_{NU,i}}}{dt}=({\omega}_{\pm{\vec{k}}_{NU,i}}%
^{\prime}+{\omega}_{\pm{\vec{k}}_{NU,i}})X_{\pm{\vec{k}}_{NU,i}}-\nonumber\\
&  {\alpha}_{G}{\omega}_{\pm{\vec{k}}_{NU,i}}Y_{\pm{\vec{k}}_{NU,i}%
}+F_{\mathrm{th}}^{\prime}+\nonumber\\
&  \frac{{\mathcal{D}}_{\pm{\vec{k}}_{NU,i}}^{(Suhl)}}{2}\left(  P_{i}{\alpha
}_{\mp{\vec{k}}_{NU,i}}^{\ast}+P_{i}^{\ast}{\alpha}_{\mp{\vec{k}}_{NU,i}%
}\right)  +\nonumber\\
&  {\mathcal{D}}_{\pm{\vec{k}}_{NU,i}}^{(SK)}\left(  {\alpha}_{\pm{\vec{k}%
}_{NU,i}}^{\ast}{\alpha}_{\pm{\vec{k}}_{NU,i}}{\alpha}_{\pm{\vec{k}}_{NU,i}%
}+\right. \nonumber\\
&  \left.  {\alpha}_{\pm{\vec{k}}_{NU,i}}{\alpha}_{\pm{\vec{k}}_{NU,i}}^{\ast
}{\alpha}_{\pm{\vec{k}}_{NU,i}}^{\ast}\right)  , \label{eq31}%
\end{align}
where ${\vec{H}}_{\mathrm{A}}$, ${\vec{H}}_{\mathrm{dc}}$, ${\vec{H}%
}_{\mathrm{th}}$ are the anisotropy, applied dc, and white thermal noise
magnetic fields, respectively. The Cartesian components ${\vec{H}%
}_{\mathrm{th}}$ are independent Gaussian fluctuators with variance $2{\alpha
}_{\mathrm{G}}k_{\mathrm{B}}T/(\gamma M_{s}V)$. $F_{\mathrm{th}}$ and
$F_{\mathrm{th}}^{\prime}$ are independent Gaussian fluctuations with variance
$2{\alpha_{\mathrm{G}}\omega}_{\pm\vec{k}_{NU,i}}\bar{n}_{\pm{\vec{k}}_{NU,i}%
}$. To summarize, the nonlinear magnon interactions induce $\vec{H}_{SW}$ in
the LLG equation governing the macrospin dynamics $\vec{m}$, while the the
latter determines $P_{i}$ that enters in finte-$k$ magnon dynamics
$(X_{\pm\vec{k}_{NU,i}},Y_{\pm\vec{k}_{NU,i}})$, resulting in a closed EOM,
that can be integrated numerically.

\subsection{\label{results1} Results and discussion}

\begin{figure}[ptb]
\includegraphics[width=0.5\textwidth]{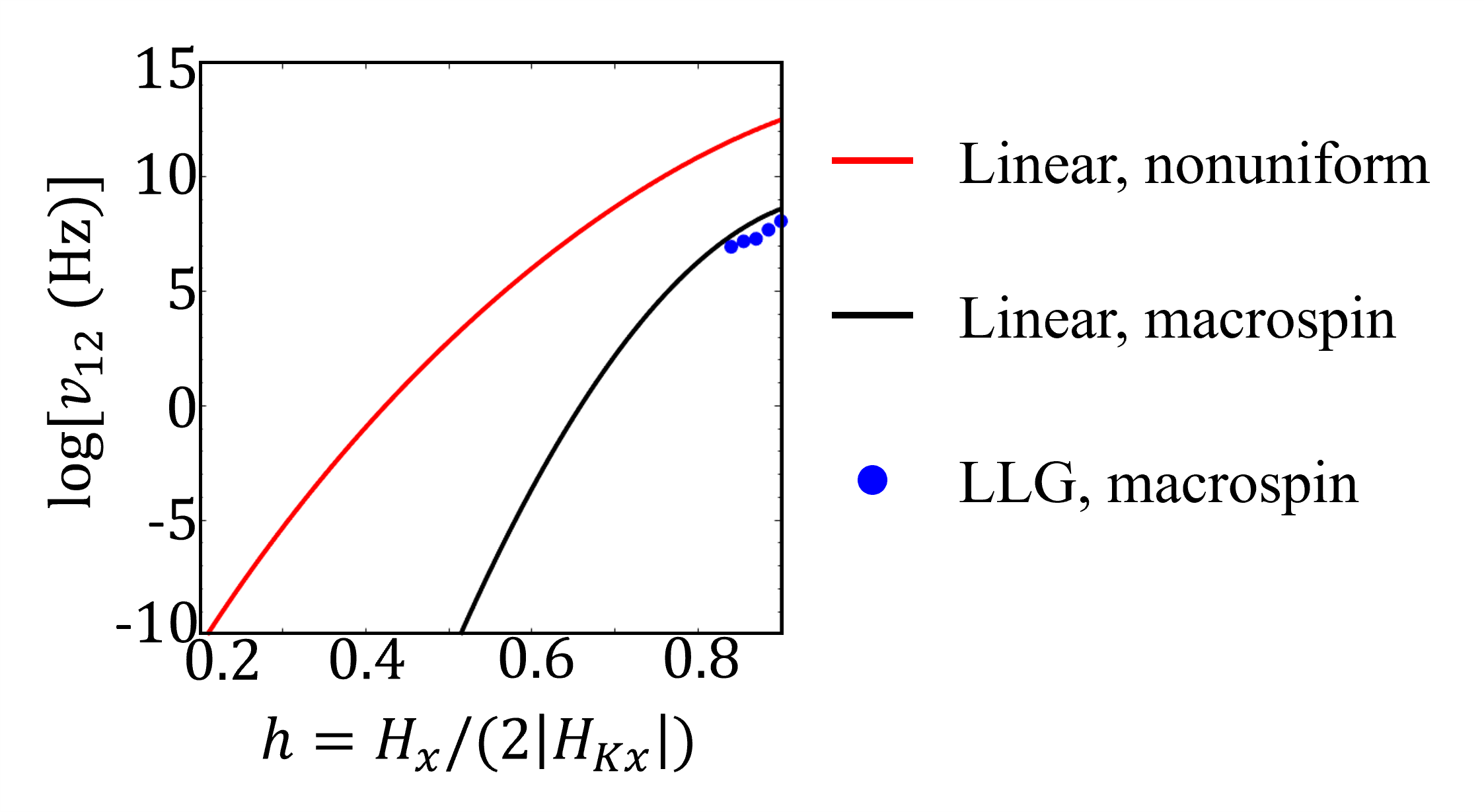}\caption{Calculated
transition frequency from the metastable to stable state, $v_{12}$, of an
in-plane magnetization as a function of an external field $H_{x}$ along the
easy axis $\hat{x}$ \textit{without} taking into account the decay of the
Kittel mode into spin waves. The easy axis in-plane anisotropy along $\hat{x}%
$, $H_{Kx}=-1.8\,$T, the out-of-plane uniaxial hard axis anisotropy
$H_{Kz}/H_{Kx}=-0.2$, $M_{s}=3.8\times10^{4}\,\mathrm{{A}/{m}}$, $L=150\,$nm,
$\mathcal{S}=\pi\times5\,\mathrm{{nm}^{2}}$, $\mathcal{A}_{ex}=5\times
10^{-12}\,\mathrm{{J}/{m}}$, $\alpha_{G}=0.01$, $\gamma=28\,\textrm{GHz}/\mathrm{T}$. The
black line is calculated from Brown's linear macrospin theory; red line is
calculated from Braun's linear nonuniform theory; blue dots are numerically
calculated from the macrospin LLG.}%
\label{fig4}%
\end{figure}

\subsubsection{\label{results1_analytic} Phenomenology}

The stochastic switching frequency $f_{s}$ is directly connected to the
curvature of the free energy or force constant at each minimum. This manifests
in the effective distribution function around the minima. When the minima and
the saddle point of the macrospin free energy are on the $x$ axis, the Kittel
mode occupation at each minimum
\begin{equation}
\langle m_{x}m_{x}\rangle=\frac{(u_{0}+v_{0})^{2}}{4}\left\{  2\mathrm{Re}%
\langle c_{0}c_{0}\rangle+2\langle n_{0}\rangle+1\right\}  , \label{eq32}%
\end{equation}
while from EOM of $\rho^{\prime}$ (see Sec. \ref{model}),
\begin{align}
&  \frac{d\langle c_{\vec{k}_{NU}}c_{-\vec{k}_{NU}}\rangle}{dt}=i\Delta
\omega_{\pm\vec{k}_{NU}}\langle c_{\vec{k}_{NU}}c_{-\vec{k}_{NU}}%
\rangle+\nonumber\\
&  i2\mathcal{D}^{(\mathrm{Suhl})}\langle c_{0}c_{0}\rangle\lbrack2\langle
n_{\pm\vec{k}_{NU}}+1\rangle]-2\xi_{\pm\vec{k}_{NU}}\langle c_{\vec{k}_{NU}%
}c_{-\vec{k}_{NU}}\rangle, \label{eq33}%
\end{align}%
\begin{align}
&  \frac{d\langle c_{0}c_{0}\rangle}{dt}=i2\mathcal{D}^{(\mathrm{Suhl}%
)}\langle c_{\vec{k}_{NU}}c_{-\vec{k}_{NU}}\rangle\lbrack2\langle
n_{0}+1\rangle]-\nonumber\\
&  2\xi_{0}\langle c_{0}c_{0}\rangle+\mathcal{F}, \label{eq34}%
\end{align}
where we introduced the phenomenological $\mathcal{F}$ as an effective force
on $\left\langle c_{0}c_{0}\right\rangle $ that triggers the stochastic
switching of $\vec{m}$. In Sec. \ref{eqb_dist}, we discussed the dependence of
$n_{0}$ on the occupation of the $\vec{k}_{NU}$ spin waves [see Eqs.
(\ref{eq15})-(\ref{eq17})]. In the steady state
\begin{align}
&  \mathrm{Re}\left\langle c_{0}c_{0}\right\rangle =\frac{\mathrm{Re}\ F}%
{2\xi_{0}+\xi_{corr}},\nonumber\\
&  \xi_{corr}=\frac{2\xi_{\pm\vec{k}_{NU}}\left(  \mathcal{D}^{(\mathrm{Suhl}%
)}\right)  ^{2}(2\langle n_{0}\rangle+1)(2\langle n_{\pm\vec{k}_{NU}}%
\rangle+1)}{\Delta\omega_{\pm\vec{k}_{NU}}^{2}+\xi_{\pm\vec{k}_{NU}}^{2}}.
\label{eq35}%
\end{align}
While it does not give a closed expression for $f_{s},$ Equation (\ref{eq35})
tells us that at a fixed $\langle n_{0}\rangle$ and $\langle n_{\pm\vec
{k}_{NU}}\rangle$, the Suhl four magnon interaction $\mathcal{D}%
^{(\mathrm{Suhl})}\neq0$, damps the correlation function $\mathrm{Re}[\langle
c_{0}c_{0}\rangle]$ by an the additional $\xi_{corr}$, and thereby reduces
$\langle m_{x}m_{x}\rangle$. In Sec. \ref{results1_calc}, we discuss the
dependence of the numerically calculated $f_{s}$ in terms of the brake on the
dynamics caused by $\xi_{corr}$.

\subsubsection{\label{results1_calc} Numerical results and discussion}

\begin{figure}[ptb]
\includegraphics[width=0.5\textwidth]{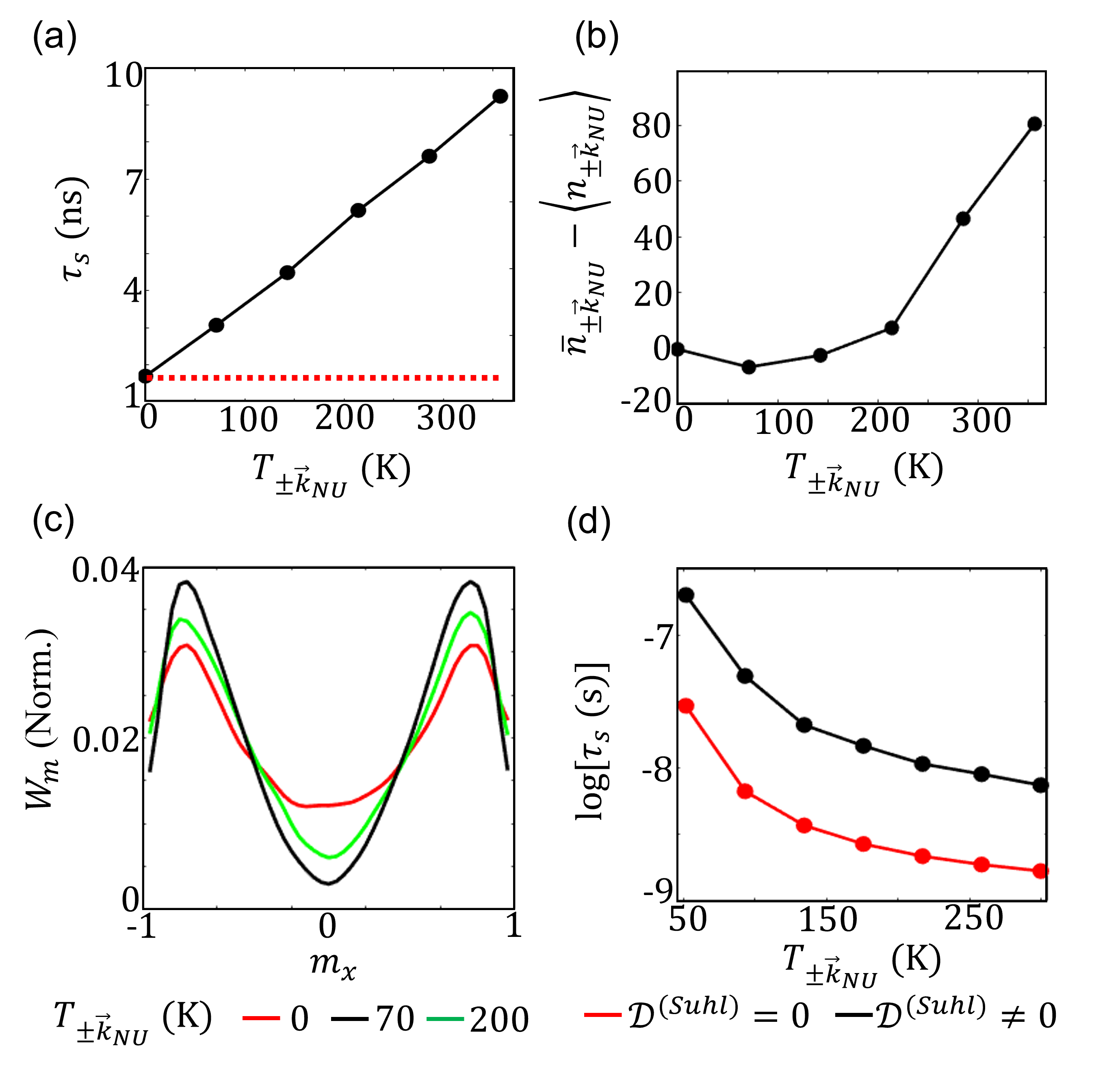}\caption{(a) $\tau
_{s}=1/f_{s}$ as a function of the magnon temperature $T_{\vec{k}_{NU}}$. The
red dashed line indicates $\tau_{s}$ when $\mathcal{D}^{(\mathrm{Suhl})}=0$.
(b) The dependence of $\bar{n}_{\vec{k}_{NU}}-\langle{n}_{\vec{k}_{NU}}%
\rangle$ on $T_{\vec{k}_{NU}}$. (c) The marginal distribution $W_{m}$ of the
magnetization along $\hat{x}$ for different $T_{\vec{k}_{NU}}$. In (a)-(c),
the macrospin environment temperature is fixed at $T_{0}=300\,$K. (d)
$\tau_{s}$ as a function of $T_{0}=T_{\pm\vec{k}_{NU}}$ without and with the
Suhl nonlinear interaction, $\mathcal{D}^{(Suhl)}=0$ (red) and $\mathcal{D}%
^{(Suhl)}\neq0$ (black), respectively. The inset shows $\tau_{r}=\tau
_{s}|_{\mathcal{D}^{(Suhl)}\neq0}/\tau_{s}|_{\mathcal{D}^{(Suhl)}=0}$. In
(a)-(d), the sample diameter $r=20\,$nm, $\theta_{s}=\pi/4$, $H_{Kx}=-75\,$mT,
$H_{Kz}=0.2\,$T, and $M_{s}=1.5\times10^{6}\,\mathrm{{A}/{m.}}$}%
\label{fig5}%
\end{figure}

\begin{figure}[ptb]
\includegraphics[width=0.5\textwidth]{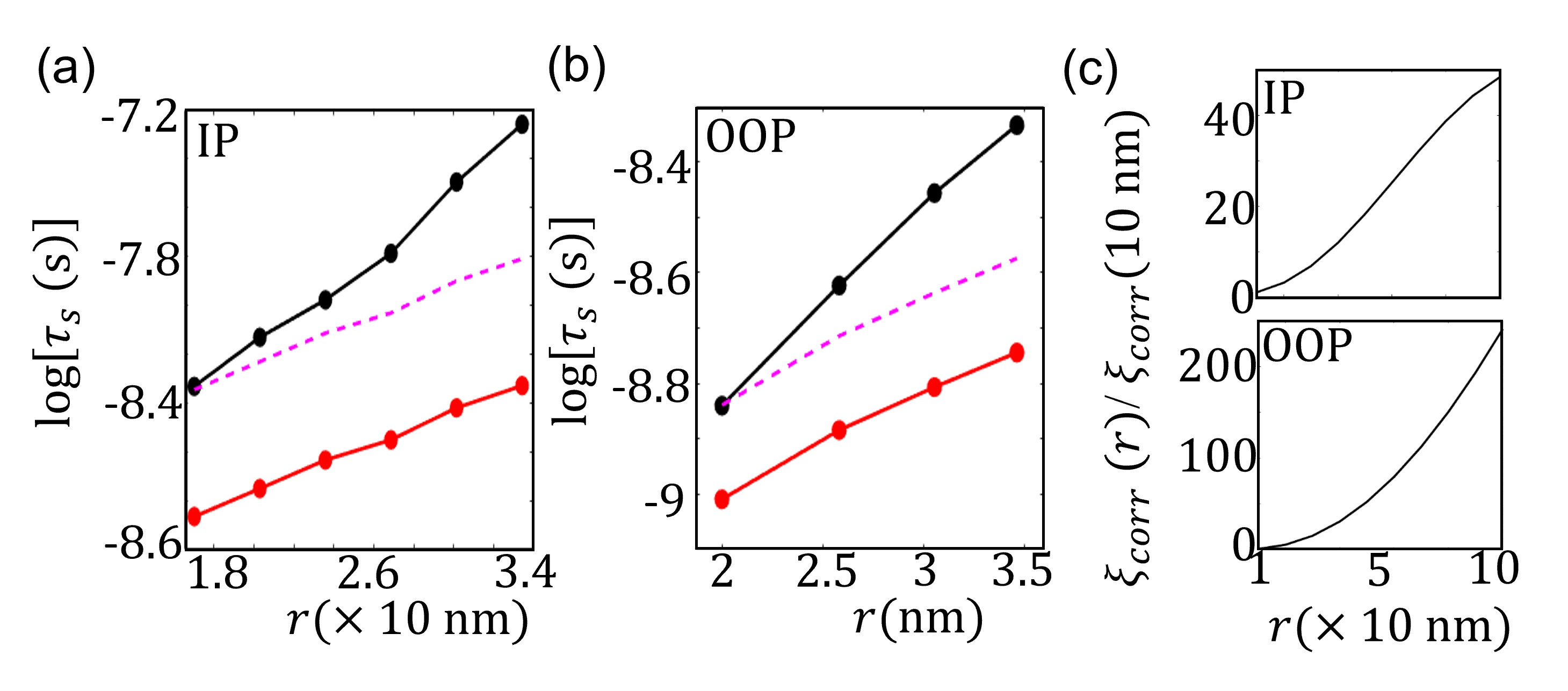}\caption{Sample radius
dependence. (a)-(b) $\tau_{s}$ as a function of $r$ when $\mathcal{D}%
^{(Suhl)}=0$ (red dots) and $\mathcal{D}^{(Suhl)}\neq0$ (black dots). The
magenta dashed lines are shifted red lines for scale comparison with the black
lines. (a) and (b) correspond to the IP and OOP cases, respectively.
$T_{0}=T_{\pm\vec{k}_{NU}}$ in (a) and (b). In (b), $H_{kx}=-50\,$mT. (c)-(d)
$\xi_{corr}$ as a function of $r$. (c) and (d) correspond to the IP and OOP
cases, respectively.}%
\label{fig6}%
\end{figure}

Here we first report numerical micromagnetic results without spin waves and
compare them with Brown's and Braun's theories, then turning to calculations
for our spin wave model.

With Fig. (\ref{fig4}) we report that $v_{12}$ from a numerical solution of
the LLG calculation for the macrospin $v_{12}^{(MS,LLG)}$ is only slightly
lower than the model $v_{12}^{(\mathrm{Brown})}$, justifying Kramers escape model.

Next, we discuss the stochastic LLG equation for the domain wall assisted
switching based on the Braun theory. We focus here on the experimentally most
interesting in-plane magnetized samples with $H_{Kx}=\mu_{0}M_{s}K_{x}%
=-1.8\,$T and a relatively weak out-of-plane hard uniaxial anisotropy,
$H_{Kz}/H_{Kx}=-0.2$. A magnetic field $H_{x}$ applied in the $\hat{x}$
direction shifts the two energy minima with respect to each other and $v_{12}$
is the transition from the high to the low state. We may then plug
$E(\theta_{1})=-H_{x}M_{s}$, $E(\theta_{sd})=H_{Kx}M_{s}(1+\epsilon^{2})$,
$\lambda_{sd,\phi}=-2H_{Kx}M_{s}(\epsilon^{2}-1)$, $\lambda_{sd,\theta
}=2H_{Kz}M_{s}(1-\epsilon^{2})$, $\lambda_{1,\phi}=-2H_{Kx}M_{s}+2H_{Kz}%
M_{s}+H_{x}M_{s}$, $\lambda_{1,\theta}=-2H_{Kx}M_{s}+H_{x}M_{s}$, and
$\epsilon=H_{x}/(2H_{Kx})$ into the Brown theory Eq. (\ref{eq23}). We use the
Braun formula simplified from Eq. (\ref{eq25}) \cite{Braun1994_1}, viz.
\begin{equation}
v_{12}\approx2\gamma|H_{Kx}|\Omega\exp^{-2\mathcal{S}\sqrt{A_{ex}|H_{Kx}%
|M_{s}}\mathcal{E}},
\end{equation}
where
\begin{equation}
\Omega=16\pi^{-3/2}\alpha_{G}L\sqrt{2\beta\mathcal{S}\sqrt{A_{ex}|H_{Kx}%
|M_{s}}}[\sqrt{Q}+\sqrt{1+Q}]^{2}\exp[-R]
\end{equation}
$\mathcal{E}=4\tanh R-4R\mathrm{sech}^{2}R$ , $R=\mathrm{{sech}}^{-1}\sqrt{h}%
$, $h=H_{x}/(2|H_{Kx}|)$, and $Q=H_{Kz}/|H_{Kx}|$.

Figure \ref{fig4} shows that $v_{12}$ as calculated by Braun's theory
$v_{12}^{(\mathrm{Braun})}$ is significantly larger than estimated by the
macrospin result $v_{12}^{(\mathrm{Brown})}$ because the saddle point of the
free energy is significantly lowered by domain walls. On the
other hand, we found only weak non-linear effects on the ratios
$v_{12}/v_{21}$ (not shown). 

Now we are ready to discuss the effect of spin waves with finite wave vector
and their interaction with the macrospin, based on our three-mode model of
Sec. \ref{model1}. We compute spin wave frequencies (i.e. detuning
$\Delta\omega_{\pm\vec{k}_{NU}}$) and coefficients $\mathcal{D}%
^{(\mathrm{Suhl})}$, $\mathcal{D}_{\vec{k}}^{(SK)}$, and $\mathcal{D}^{(CK)}$
for the ultrathin films with $d=2\,$nm \cite{Krivosik2010} used in the MTJ
experiments \cite{Borders2019,Hayakawa2021,Kanai2023,Kaneko2024}. The
dispersion of coefficients and frequencies are evaluated for the lowest
frequency in-plane standing waves with $\vec{k}_{NU,+(-)}=(\pi/r)\vec
{m}_{1(2)}$ [see Fig. \ref{fig1}(a)], where $r$ is the radius of the sample.
For the IP magnetized sample, at a certain $|\vec{k}|$, $\omega_{\vec{k}}$ and
interaction coefficients are functions of the in-plane angle between the
magnetization and $\vec{k}$, $\theta_{\vec{k}}$. The detuning $\Delta
\omega_{\pm\vec{k}_{NU}}$ and interaction coefficients are smallest and
strongest, respectively, for $\theta_{\vec{k}}=0$. Therefore, for the IP
configuration, we choose $\vec{k}_{NU,+(-)}$ with $\theta_{\vec{k}_{NU,+(-)}%
}=0$. For the OOP magnetized sample, $\omega_{\vec{k}}$ and interaction
coefficients are isotropic for in-plane $\vec{k}$, so the choice of in-plane
angle for $\vec{k}_{NU,+(-)}$ does not matter. In the elliptical samples of
Ref. \cite{Kanai2023}, an IP easy axis leads to a $\theta_{s}$-dependence of
$|\vec{k}_{NU}|$ with changing $\theta_{s}$ that we disregard here. Therefore,
for the IP magnetized sample
\begin{align}
&  \Delta\omega_{\pm\vec{k}_{NU}}\approx\omega_{M}\sqrt{|H_{Ke}H_{Kh}%
|} \frac{A_{ex}k_{NU}^{2}}{2|H_{Ke}\cos\theta_{s}|},\nonumber\\
&  \mathcal{D}^{(Suhl)}\approx\frac{\omega_{M}}{2}\left\{  \left[
H_{Kh}+A_{ex}k_{NU}^{2}+H_{Ke}\times\right.  \right. \nonumber\\
&  \left.  \left.  (\sin^{2}\theta_{s}-2\cos^{2}\theta_{s})\right]  u_{0}%
^{2}+\right. \nonumber\\
&  \left.  \left[  3(H_{Ke}\sin^{2}\theta_{s}-H_{Kh})\right]  u_{0}%
v_{0}\right\}  ,\nonumber\\
&  u_{0}\approx\sqrt{\frac{1}{4}\sqrt{\frac{H_{Kh}}{|H_{Ke}|\cos^{2}\theta_{s}}}+\frac{1}{2}},\nonumber\\
&  v_{0}\approx\sqrt{\frac{1}{4}\sqrt{\frac{H_{Kh}}{|H_{Ke}|\cos^{2}
\theta_{s}}}-\frac{1}{2}}. \label{eq36}%
\end{align}
For the OOP samples,
\begin{align}
&  \Delta\omega_{\pm\vec{k}_{NU}}\approx\omega_{M}|H_{Ke}|A_{ex}k_{NU}^{2},\nonumber\\
&  \mathcal{D}^{(Suhl)}\approx\frac{\omega_{M}}{2}\times \left(
-2|H_{Ke}|+2A_{ex}%
k_{NU}^{2}\right) u_{0}^{2}  ,\nonumber\\
&  u_{0}=1,v_{0}=0. \label{eq37}%
\end{align}

In the following we focus on the IP case, but the main conclusions hold for
the OOP samples. First, we explore the dependence of $\tau_{s}=1/f_{s}$ on the
spin-wave temperature $T_{\pm\vec{k}_{NU,i}}=T_{\pm\vec{k}_{NU}}$ for constant
macrospin reservoir temperature $T_{0}=300$\ K constant at . According to
Figure \ref{fig5}(a) $\tau_{s}$ monotonically increases with $T_{\pm\vec
{k}_{NU}}$ as expected since $\xi_{corr}\propto\langle n_{\pm\vec{k}_{NU}%
}\rangle\propto T_{\pm\vec{k}_{NU}}$. When $T_{\pm\vec{k}_{NU}}\rightarrow0$
there are no $\vec{k}_{NU}$ magnons and $\xi_{corr}\rightarrow0$, i.e.
$\tau_{s}$ becomes that when $\mathcal{D}^{(Suhl)}\rightarrow0$ [see the red
dotted line in Fig. \ref{fig5}(a)]. Figure \ref{fig5}(b) shows that $\bar
{n}_{\pm\vec{k}_{NU}}-\langle{n}_{\pm\vec{k}_{NU}}\rangle<0$ for small
$T_{\pm\vec{k}_{NU}}$, and becomes positive with increasing $T_{\pm\vec
{k}_{NU}}$. This is in line with our analysis in Sec. \ref{eqb_dist} of the
balance between the effective temperatures of the Kittel mode and magnon pair
[see Fig. \ref{fig1}(c) bottom panel]. The probability distribution $W_{m_{x}%
}$ of the magnetization along the $\hat{x}$ direction in Figure \ref{fig5}(c)
shows an increasing localization at the energy minima with increasing
$T_{\pm\vec{k}_{NU}}$, reflecting the increased damping of the fluctuations
expected from Sec. \ref{results1_analytic}.

Next, we study the dependence on $T_{0}=T_{\pm\vec{k}_{NU}}=T$. According to
Figure \ref{fig4}(d) $\tau_{s}$ is always enhanced when the Suhl interaction
$\mathcal{D}^{(\mathrm{Suhl})}\neq0$. The ratio
\begin{equation}
\tau_{r}=\frac{\tau_{s}|_{\mathcal{D}^{(\mathrm{Suhl})}\neq0}}{\tau
_{s}|_{\mathcal{D}^{(\mathrm{Suhl})}=0}}\gg1
\end{equation}
does not depend strongly on $T$ which proves that spin waves mainly affect the
attempt frequency $\tau_{0}$, in agreement with experiments \cite{Kanai2023}.
In the experiments by Kanai et al. \cite{Kanai2023}, we extract $\tau_{0}$ by
measuring $f_{s}$ from resistance fluctuations in tunneling
magneto-resistance, as a function of $\theta_{s}$. The latter is varied by
changing the magnetic field along $\hat{y}$ [see Fig. \ref{fig3}(a), top
panel], leading to $\tau_{0}$ estimations an order of magnitude larger than
what is expected from macrospin LLG calculations or Brown theory.

From an application point of view it is of great interest to reduce the
switching time by changing the sample dimensions \cite{Jinnai2020}. Figures
\ref{fig6}(a) and (b) show that $\tau_{r}$ indeed decreases with the sample
radius $r$ for both the IP and OOP samples. $\xi_{corr}$ [see Eq.
(\ref{eq35})] in Fig. \ref{fig6}(c) monotonically increases with $r$, since
$\Delta\omega_{\pm\vec{k}_{NU}}\gg\xi_{\pm\vec{k}_{NU}}$, $\Delta\omega
_{\pm\vec{k}_{NU}}\propto1/r^{2}$, and $\xi_{\pm\vec{k}_{NU}}=\alpha_{G}%
\omega_{\pm\vec{k}_{NU}}\propto(a+b/r^{2})$, $\langle n_{\pm\vec{k}_{NU}%
}\rangle\propto1/\omega_{\pm\vec{k}_{NU}}$, and $\mathcal{D}^{(\mathrm{Suhl}%
)}\propto(a^{\prime}+b^{\prime2})$, where $a$, $a^{\prime}$, $b$ and
$b^{\prime}$ are constants related to material parameters [see Eqs.
(\ref{eq36})-(\ref{eq37})]. This dependence agrees with recent observations
\cite{Kaneko2024} of an increase of $\tau_{0}$ with $r$ for OOP samples in
contrast to the N\'{e}el-Brown theory that predicts a decrease in $\tau_{0}$ with
magnetic volume [see Eq. (\ref{eq22})].

\section{Conclusion}

We theoretically study the interactions of uniform magnetization precession
with finite-momentum spin waves at finite temperatures on fluctuations around
the equilibrium magnetizations and the stochastic switching dynamics of the
magnetization between two equilibrium directions. We analytically and
numerically show that a four-magnon interaction attracts magnon numbers of the
Kittel mode and coupled spin waves, while suppressing the random switching of the magnetization.
The temperature affects the stochastic switching frequency dominantly via the
attempt frequency that decreases with the effective temperature of the
interacting finite-$k$ spin waves. Shrinking the sample for a fixed film thickness
increases the attempt frequency. The performance of MTJs in applications such
as probabilistic computing can therefore be boosted by shrinking the sample
dimensions and selective cooling of spin waves with finite momenta. Future
theories and experiments should be designed to explore the active cooling.
Theoretically, more in-depth analytical and numerical models should delineate
the intricacies of nonlinear interactions as well as domain wall nucleation
and motion that may cause numerical corrections to switching frequencies
computed here.

\acknowledgments We acknowledge support by JSPS KAKENHI (Grants Nos. 21K13847,
19H00645, 22H04965, 20H02178, 24H02231, 24H02235, 24H00039), JST CREST (Grant No.
JPMJCR19K3), JST-ASPIRE (Grant No. JPMJAP2322), JST PRESTO (Grant No. JPMJPR21B2), MEXT X-NICS JPJ011438,
Shimadzu Science Foundation, Takano Research Foundation, and Cooperative
Research Projects of RIEC.


\begin{thebibliography}{99}                                                                                               %


\bibitem {Neel1949}L. N\'eel, \textit{Theorie du trainage magnetique des
ferromagnetiques en grains fins avec application aux terres cuites}, Ann.
Geophys. \textbf{5}, 99 (1949).

\bibitem {Brown1963}W. F. Brown, \textit{Thermal Fluctuations of a
Single-Domain Particle}, Phys. Rev. \textbf{130}, 1677 (1963).

\bibitem {Brown1979}W. F. Brown, \textit{Thermal Fluctuations of Fine
Ferromagnetic Particles}, IEEE Trans. Mag. \textbf{15}, 1196 (1979).

\bibitem {Hayakawa2021}K. Hayakawa, S. Kanai, T. Funatsu, J. Igarashi, B.
Jinnai, W. A. Borders, H. Ohno, and S. Fukami, \textit{Nanosecond Random
Telegraph Noise in In-Plane Magnetic Tunnel Junctions}, Phys. Rev. Lett.
\textbf{126}, 117202 (2021).

\bibitem {Safranski2021}C. Safranski, J. Kaiser, P. Trouilloud, P. Hashemi, G.
Hu, and J. Z. Sun, \textit{Demonstration of Nanosecond Operation in Stochastic
Magnetic Tunnel Junctions}, Nano. Lett. \textbf{21}, 2040-2045 (2021).

\bibitem {Chudnovsky1988}E. M. Chudnovsky and L. Gunther, \textit{Quantum
Tunneling of Magnetization in Small Ferromagnetic Particles}, Phys. Rev. Lett.
\textbf{60}, 661 (1988).

\bibitem {Awschalom1992}D. D. Awschalom, J. F. Smyth, G. Grinstein, D. P.
DiVincenzo, and D. Loss, \textit{Macroscopic Quantum Tunneling in Magnetic
Proteins}, Phys. Rev. Lett. \textbf{68}, 3092 (1992).

\bibitem {Barkhausen1919}H. Barkhausen, Phys. Z. \textbf{20}, 401 (1919).

\bibitem {Spasojevic1996}Djordje Spasojevic, Srdjan Bukvic, Sava Milosevic,
and H. Eugene Stanley, \textit{Barkhausen noise: Elementary signals, power
laws, and scaling relations}, Phys. Rev. E \textbf{54}, 2531 (1996).

\bibitem {Lachance2020}D. Lachance-Quirion, S. P. Wolski, Y. Tabuchi, S. Kono,
K. Usami, and Y. Nakamura, \textit{Entanglement-based single-shot detection of
a single magnon with a superconducting qubit}, Science \textbf{367}, 425 (2020).

\bibitem {Elyasi2020}M. Elyasi, Y. M. Blanter, G. E. W. Bauer,
\textit{Resources of nonlinear cavity magnonics for quantum information},
Phys. Rev. B \textbf{101}, 054402 (2020).

\bibitem {Yuan2022}H. Yuan, Y. Cao, A. Kamra, R. A. Duine, and P. Yan,
\textit{Quantum magnonics: When magnon spintronics meets quantum information
science,} Phys. Rep. \textbf{965}, 1 (2022).

\bibitem {Rameshti2022}B. Z. Rameshti, S. V. Kusminskiy, J. A. Haigh, K.
Usami, D. Lachance-Quirion, Y. Nakamura, C.-M. Hu, H. X. Tang, G. E. W. Bauer,
and Y. M. Blanter, \textit{Cavity magnonics}, Phys. Rep. \textbf{979}, 1 (2022)

\bibitem {Chumak2022}A. Chumak et al., \textit{Advances in magnetics roadmap
on spin-wave computing}, IEEE Trans. Magn. \textbf{58}, 1 (2022).

\bibitem {Hioki2021}T. Hioki, H. Shimizu, T. Makiuchi, and E. Saitoh,
\textit{State tomography for magnetization dynamics}, Phys. Rev. B
\textbf{104}, L100419 (2021).

\bibitem {Makiuchi2024}T. Makiuchi, T. Hioki, H. Shimizu, K. Hoshi, M. Elyasi,
K. Yamamoto, N. Yokoi, A. Serga, B. Hillebrands, G. E. W. Bauer, and E.
Saitoh, \textit{Persistent magnetic coherence in magnets}, Nat. Mater. (2024).

\bibitem {Makiuchi2021}T. Makiuchi, T. Hioki, Y. Shimazu, Y. Oikawa, N. Yokoi,
S. Daimon, and E. Saitoh, \textit{Parametron on magnetic dot: Stable and
stochastic operation}, Appl. Phys. Lett. \textbf{118}, 022402 (2021)

\bibitem {Elyasi2022}M. Elyasi, E. Saitoh, and G. E. W. Bauer,
\textit{Stochasticity of the magnon parametron}, Phys. Rev. B \textbf{105},
054403 (2022).

\bibitem {Endean2014}D. E. Endean, C. T. Weigelt, R. H. Victora, and E. Dan
Dahlberg, \textit{Tunable random telegraph noise in individual square
permalloy dots}, Appl. Phys. Lett. \textbf{104}, 252408 (2014).

\bibitem {Talatchian2021}P. Talatchian, M. W. Daniels, A. Madhavan, M. R.
Pufall, E. Jue, W. H. Rippard, J. J. McClelland, and M. D. Stiles,
\textit{Mutual control of stochastic switching for two electrically coupled
superparamagnetic tunnel junctions}, Phys. Rev. B \textbf{104}, 054427 (2021).


\bibitem {Camsari2017}K. Y. Camsari, R. Faria, B. M. Sutton, and S. Datta,
\textit{Stochastic p-Bits for Invertible Logic}, Phys. Rev. X \textbf{7},
031014 (2017).

\bibitem {Borders2019}W. A. Borders, A. Z. Pervaiz, S. Fukami, K. Y. Camsari,
H. Ohno, and S. Datta, \textit{Integer factorization using stochastic magnetic
tunnel junctions}, Nature \textbf{573}, 390-393 (2019).

\bibitem {Vodenicarevic2017}D. Vodenicarevic, N. Locatelli, A. Mizrahi, J. S.
Friedman, A. F. Vincent, M. Romera, A. Fukushima, K. Yakushiji, H. Kubota, S.
Yuasa, S. Tiwari, J. Grollier, and D. Querlioz, \textit{Low-Energy Truly
Random Number Generation with Superparamagnetic Tunnel Junctions for
Unconventional Computing}, Phys. Rev. Appl. \textbf{8}, 54045 (2017).

\bibitem {Mizrahi2018}A. Mizrahi, T. Hirtzlin, A. Fukushima, H. Kubota, S.
Yuasa, J. Grollier, and D. Querlioz, \textit{Neural-like computing with
populations of superparamagnetic basis functions}, Nat. Commun. \textbf{9},
1533 (2018).

\bibitem {Kaiser2022}J. Kaiser, W. A. Borders, K. Y. Camsari, S. Fukami, H.
Ohno, and S. Datta, \textit{Hardware-Aware In Situ Learning Based on
Stochastic Magnetic Tunnel junctions}, Phys. Rev. Appl. \textbf{17}, 014016 (2022).

\bibitem {Singh2024}N. S. Singh, K. Kobayashi, Q. Cao, K. Selcuk,
T. Hu, S. Niazi, N. A. Aadit, S. Kanai, H. Ohno, S. Fukami, K. Y. Camsari, \textit{CMOS plus stochastic nanomagnets enabling
heterogeneous computers for probabilistic inference and learning}, Nat. Commun. \textbf{15}, 2685 (2024). 

\bibitem {Si2024}J. Si, S. Yang, Y. Cen, J. Chen, Y. Huang, Z. Yao,
D.-J. Kim, K. Cai, J. Yoo, X. Fong, H. Yang, \textit{Energy-efficient superparamagnetic Ising machine and its application to traveling salesman problems}, Nat. Commun. \textbf{15}, 3457 (2024). 

\bibitem {Ikeda2010}S. Ikeda, K. Miura, H. Yamamoto, K. Mizunuma, H. D. Gan,
M. Endo, S. Kanai, J. Hayakawa, F. Matsukura, H. Ohno, \textit{A
perpendicular-anisotropy CoFeB-MgO magnetic tunnel junction}, Nat Mater
\textbf{9}, 721 (2010).

\bibitem {Watanabe2018}K. Watanabe, B. Jinnai, S. Fukami, H. Sato, and H.
Ohno, \textit{Shape anisotropy revisited in single-digit nanometer magnetic
tunnel junctions}, Nat. Commun. \textbf{9}, 663 (2018).

\bibitem {Jinnai2020}B. Jinnai, K. Watanabe, S. Fukami, and H. Ohno,
\textit{Scaling magnetic tunnel junction down to single-digit
nanometers-Challenges and prospects}, Appl. Phys. Lett. \textbf{116}, 160501 (2020).

\bibitem {Funatsu2022}T. Funatsu, S. Kanai, J. Ieda, S. Fukami, and H. Ohno,
\textit{Local bifurcation with spin-transfer torque in superparamagnetic
tunnel junctions}, Nat. Commun. \textbf{13}, 4079 (2022).

\bibitem {Kramers1940}H. A. Kramers, \textit{Brownian motion in a field of
force and the diffusion model of chemical reactions}, Physica \textbf{7}, 284 (1940).

\bibitem {Braun1993}H.-B. Braun, \textit{Thermally Activated Magnetization
Reversal in Elongated Ferromagnetic Particles}, Phys. Rev. Lett. \textbf{71} ,
3557 (1993).

\bibitem {Braun1994}H.-B. Braun, \textit{Kramers rate theory, broken
symmetries, and magnetization reversal (invited)}, J. Appl. Phys. \textbf{76},
6310 (1994).

\bibitem {Braun1994_1}H.-B. Braun, \textit{Statistical mechanics of nonuniform
magnetization reversal}, Phys. Rev. B \textbf{50}, 16501 (1994).

\bibitem {Braun1994_2}H.-B. Braun, \textit{Fluctuations and instabilities of
ferromagnetic domain-wall pairs in an external magnetic field}, Phys. Rev. B
\textbf{50}, 16485 (1994).

\bibitem {Krause2009}S. Krause, G. Herzog, T. Stapelfeldt, L. Berbil-Bautista,
M. Bode, E.Y. Vedmedenko, and R. Wiesendanger, \textit{Magnetization Reversal
of Nanoscale Islands: How Size and Shape Affect the Arrhenius Prefactor},
Phys. Rev. Lett. \textbf{103}, 127202 (2009).

\bibitem {Bessarab2013}P. F. Bessarab, V. M. Uzdin, and H. Jonsson,
\textit{Size and Shape Dependence of Thermal Spin Transitions in Nanoislands},
Phys. Rev. Lett. \textbf{110}, 020604 (2013).

\bibitem {Sala2023}G. Sala, J. Meyer, A. Flechsig, L. Gabriel, and P.
Gambardella, \textit{Deterministic and stochastic aspects of current-induced
magnetization reversal in perpendicular nanomagnets}, Phys. Rev. B
\textbf{107}, 214447 (2023).

\bibitem {Kanai2021}S. Kanai, K. Hayakawa, H. Ohno, and S. Fukami,
\textit{Theory of relaxation time of stochastic nanomagnets}, Phys. Rev. B
\textbf{103}, 094423 (2021).

\bibitem {Arrhenius1889}S. Arrhenius, \textit{\"{U}ber die
Reaktionsgeschwindigkeit bei der Inversion von Rohrzucker durch
S\"{a}uren}, Zeitschrift f\"{u}r Physikalische Chemie. \textbf{4U}, 226 (1889).

\bibitem {Kanai2023}S. Kanai, K. Hayakawa, M. Elyasi, K. Kobayashi, J.
Igarashi, B. Jinnai, W. A. Borders, G. E. W. Bauer, H. Ohno, and S. Fukami1,
\textit{Stochastic switching time constant and instability in nanomagnets}, submitted.

\bibitem {Kaneko2024}H. Kaneko, R. Ota, K. Kobayashi, S. Kanai, M. Elyasi, G.
E. W. Bauer, H. Ohno, and S. Fukami, \textit{Temperature dependence of the
properties of stochastic magnetic tunnel junction with perpendicular
magnetization}, APEX \textbf{17}, 053001 (2024).

\bibitem {Aspelmeyer2014}M. Aspelmeyer, T. J. Kippenberg, and F. Marquardt,
\textit{Cavity optomechanics}, Rev. Mod. Phys. \textbf{86}, 1391 (2014).

\bibitem {Krivosik2010}P. Krivosik and C. E. Patton, \textit{Hamiltonian
formulation of nonlinear spin-wave dynamics: Theory and applications}, Phys.
Rev. B \textbf{82}, 184428 (2010).

\bibitem {Carmichael1}H. J. Carmichael, \textit{Statistical Methods in Quantum
Optics 1}, Springer (1999).

\bibitem {Walls2008}D. F. Walls and G. J. Milburn, \textit{Quantum Optics},
Springer (2008).

\bibitem {Suhl1957}H. Suhl, \textit{The Theory of Ferromagnetic Resonance at
High Signal Powers}, Phys. Chem. Solids \textbf{1}, 209 (1957).

\bibitem {Lvov1994}V. S. Lvov, \textit{Wave Turbulence Under Parametric
Excitation}, (Springer-Verlag, 1994).

\bibitem {Rezende2020}S. M. Rezende, \textit{Fundamentals of magnonics},
Springer, Berlin, (2020).

\bibitem {Kurebayashi2011}H. Kurebayashi, O. Dzyapko, V. E. Demidov, D. Fang,
A. J. Ferguson, and S. O. Demokritov, \textit{Controlled enhancement of
spin-current emission by three-magnon splitting}, Nat. Mater. \textbf{10}, 660 (2011).

\bibitem {Lee2023}O. Lee, K. Yamamoto, M. Umeda, C. W. Zollitsch, M. Elyasi,
T. Kikkawa, E. Saitoh, G. E. W. Bauer, H. Kurebayashi, \textit{Nonlinear
magnon polaritons}, Phys. Rev. Lett. \textbf{130}, 046703 (2023).

\bibitem {Sheng2023}L. Sheng, M. Elyasi, J. Chen, W. He, Y. Wang, H. Wang, H.
Feng, Y. Zhang, I. Medlej, S. Liu, W. Jiang, X. Han, D. Yu, J.-P. Ansermet, G.
E. W. Bauer, H. Yu, \textit{Nonlocal Detection of Interlayer Three-Magnon
Coupling}, Phys. Rev. Lett. \textbf{130}, 046701 (2023).

\bibitem {Carmichael2}H. J. Carmichael, \textit{Statistical Methods in Quantum
Optics 2}, Springer (2008).

\bibitem {Langer1969}J.-S. Langer, \textit{Statistical Theory of the Decay of
Metastable States}, Ann. Phys. \textbf{54}, 258 (1969).

\bibitem {Landauer1961}R. Landauer and J. A. Swanson, \textit{Frequency
Factors in the Thermally Activated Process},Phys. Rev. \textbf{121}, 1668 (1961).

\bibitem {Kinsler1991}P. Kinsler and P. D. Drummond, \textit{Quantum dynamics
of the parametric oscillator}, Phys. Rev. A \textbf{43}, 6194 (1991).

\end{thebibliography}
\end{document}